\documentclass[a4paper,fleqn,usenatbib,unicode]{mnras}

\usepackage[T1]{fontenc}
\usepackage{ae,aecompl}

\usepackage{graphicx}	
\usepackage{amsmath}	
\usepackage{amssymb}	
\usepackage{commath}
\usepackage[utf8x]{inputenc}
\usepackage{textgreek}
\usepackage{tikz}
\usepackage{multirow}
\usepackage{bm}

\title[Fragmentation of self-gravitating discs]{A two-step gravitational cascade
for the fragmentation of self-gravitating discs}

\author[N. Brucy and P. Hennebelle]{
Noé Brucy,$^{1}$\thanks{e-mail: noe.brucy@cea.fr}
Patrick Hennebelle$^{1}$
\\
$^{1}$AIM, CEA, CNRS, Université Paris-Saclay, Université de Paris, Sorbonne Paris Cité, F-91191 Gif-sur-Yvette, France\\
}

\date{Accepted XXX. Received YYY; in original form ZZZ}
\makeatletter
\def\input@path{{./}{figures/}{..}{../figures}}
\makeatother
\graphicspath{{./}{figures/}{..}{../figures}}
\pubyear{2020}

\begin{document}
\tikzset{every picture/.style={line width=0.75pt}} 

\label{firstpage}
\pagerange{\pageref{firstpage}--\pageref{lastpage}}
\maketitle

\begin{abstract}
Self-gravitating discs are believed to play an important role in astrophysics in particular regarding the star and planet formation process. 
In this context, discs subject to an idealized cooling process, characterized by a cooling time-scale β expressed in unit of orbital timescale, 
have been extensively studied. We take advantage of the Riemann solver and the 3D Godunov scheme implemented in the code Ramses 
to perform high-resolution simulations, complementing previous studies that have used smoothed particle hydrodynamics (SPH) or 2D grid codes.
We observe that the critical value of β for which the disc fragments is consistent with most previous results, and is not well converged with resolution.
By studying the probability density function of the fluctuations of the column density (Σ-PDF), we argue that there is no strict separation between the fragmented and the unfragmented regimes but rather a smooth transition with the probability of apparition of fragments steadily diminishing as the cooling becomes less effective. 
We find that the high column density part of the Σ-PDF follows a simple power-law whose slope turns out to be proportional to β and we propose an explanation 
based on  the balance between cooling and heating through gravitational stress. Our explanation suggests that a more efficient cooling requires more heating 
implying a larger fraction of dense material which, in the absence of characteristic scales, results in a shallower scale-free power-law.
We  propose that the gravitational cascade proceeds in two steps, first the formation of a dense filamentary spiral pattern through a sequence of quasi-static equilibrium triggered by the viscous transport of angular momentum, and second the collapse alongside these filaments that 
eventually results in the formation of bounded fragments. 
\end{abstract}




\section{Introduction}
\label{sec:intro}

\begin{table}
	\caption{Main notations and abbreviations used in the article.}
  \begin{center}
	\begin{tabular}{ll}
	\hline\noalign{\smallskip}
	Notation &  Description \\
	\noalign{\smallskip}
	\hline
	\noalign{\smallskip}
	$G$                       & Gravitational constant \\
	$M_\star$                 & Mass of the central object  \\ 
	$\bm{g_\star}$                 & Gravity field due to the star \\
	$\bm{g_\mathrm{gas}}$          & Gravity field due to the gas \\
	$\rho$                    & Gas density   \\
	$\Sigma$                  & Column density \\
	$\Omega$                  & Rotation frequency \\
	$\alpha$                  &  Shakura and Sunyaev $\alpha$ parameter\\
    $\alpha_{\mathrm{grav}}$  &  Gravitational stress contribution to $\alpha$\\
    $h$                       & Scale height of the disc\\
	$Q $                      & Toomre's parameter \\
	$\kappa$                  & Epicyclic frequency \\
	$M_d$                     & Disc's mass \\
	$r_d$                     & Disc's maximal radius  \\
    $t_{\text{cool}}$         & Cooling time \\
    $\beta = \Omega t_{\text{cool}}$ & Cooling parameter\\
    $\beta_{\text{crit}}$     & Critical value of $\beta$ for fragmentation\\
    $l$                       & Refinement level\\
    $l_{\max}$                & Maximal refinement level\\
    $L$                       & Size of the simulation box (code unit) \\
    $\bm{v}$                  & Gas velocity   \\
    $v_{\mathrm{kepl}}$       & Keplerian speed   \\
    $P$                       & Gas pressure   \\
    $\gamma$                  & Adiabatic index  \\
    $T$                       & Gas temperature   \\
    $E$                       & Gas total energy   \\
    $U$                       & Gas internal energy   \\
    $c_s$                     & Sound speed  \\
    \noalign{\smallskip}
    \hline
	\noalign{\smallskip}
    ORP                       & Outer Rotation Period\\
	FSP                       & Filamentary Spiral Pattern\\
    \noalign{\smallskip}
    \hline
	\noalign{\smallskip}
	$\sigma = \Sigma / \overline{\Sigma}$  & Fluctuation of column density \\
	$\mathcal{P}_\beta$       & PDF of $\log(\sigma)$ for a given value of $\beta$ \\
    $\sigma_0$                & See figure \ref{fig:sigma0_P0} \\
    $P_0$                     & See figure \ref{fig:sigma0_P0} \\
    $s$                       & Power-law slope of $\mathcal{P}_\beta$ \\
    \noalign{\smallskip}
    \hline
	\noalign{\smallskip}
    $g_\mathrm{fil}$          & Gravity field toward the centre of filament \\
    $R_\mathrm{fil}$          & Rotational support within filaments \\
    $f_{P, \mathrm{fil}}$     & Pressure force within filaments \\

	\noalign{\smallskip}
	\hline
	\noalign{\smallskip}
	
	$\bm{V}$                  & Vector \\
	$ V_a$                    & The $a$ component of vector $\bm{V}$ \\
	$\overline{X}$            & Azimuthal mean of $X$ \\
	$\bm{e_a}$                & Unitary vector for the axis $a$\\

	\noalign{\smallskip}
	\hline
	\end{tabular} 
  \end{center}
	\label{tbl:notation}
\end{table}

Discs are ubiquitous in astrophysics and a key property is their ability to fragment. 
Far enough from the centre, the gas can collapse under its own gravity and forms bound objects. This is of prime importance in protoplanetary disc around young stars for instance, since it is a possible scenario for planet formation \citep{1998Natur.393..141B, 2000ApJ...536L.101B}.
Under which conditions this collapse can occur is still poorly understood.
First, the disc must be gravitationally unstable. This was quantified by
 \cite{1964ApJ...139.1217T} with the so-called Toomre $Q$ parameter
 \begin{equation}
\label{eq:toomre}
Q = \dfrac{c_s \kappa}{\pi G \Sigma}.
\end{equation}
This parameter is obtained by a linear stability analysis. It takes into account the competing effects of the self-gravitation (via the column density $\Sigma$ and the gravitational constant $G$) versus the thermal support (via the sound speed $c_s$) and the tidal shear (via the epicyclic frequency $\kappa$, equals to the rotation rate $\Omega$ for Keplerian rotation).
Values of $Q \lesssim 1$ correspond to unstable discs.

Since gravitational instabilities result in the heating of the disc, the cooling rate must be accounted for.
\cite{2001ApJ...553..174G} ran two-dimensional shearing-box simulations of a local part of a thin disc. He used a simple cooling model, removing internal energy with a cooling time 
\begin{equation}
t_{\text{cool}} = \beta \Omega^{-1},
\end{equation}
where $\beta$ is a free parameter. The author has shown that if $\beta~\gtrsim~\beta_{\text{crit}}~=~3$, then the disc eventually reach a steady state where the cooling is balanced by the heating created by the dissipation of the turbulence generated by the gravitational instability. In \cite{1973A&A....24..337S}'s $\alpha$ formalism, this balance writes 
\begin{equation}
\label{eq:alphabeta}
\alpha = \dfrac{4}{9\gamma(\gamma -1)\beta},
\end{equation}
where $\gamma$ is the adiabatic index.
On the contrary, if the cooling is stronger and $\beta~\lesssim~\beta_{\text{crit}}$, then the disc fragments.

Gammie's results were extended to three-dimensional (3D) global discs by \cite{2003MNRAS.339.1025R} using smoothed particle dynamics (SPH) simulations.
\cite{2004MNRAS.351..630L,2005MNRAS.358.1489L} found that the dissipation of energy in steady-state discs was well described by a local viscous approach. \cite{2005MNRAS.364L..56R} found a value $\beta_\mathrm{crit}$ between 6 and 7 for $\gamma = 5/3$ for discs ten times less massive than the star. Using Equation~\eqref{eq:alphabeta}, they interpreted this limit as a maximum value of the stress $\alpha$ a disc can undergo before fragmenting, and estimate it as $\alpha_{\max} \sim 0.06$.
Interestingly, \cite{2011MNRAS.410..559M} found that the value of $\beta_\mathrm{crit}$ found by \cite{2005MNRAS.364L..56R} is a function of the local disc mass, and \cite{2011MNRAS.411L...1M} found that the value was not converged with resolution.
Several authors tried to explain this non-convergence. \cite{2011MNRAS.413.2735L} argue that the resolution requirements were not fulfilled in all previous SPH simulations. On the other hand, \cite{2012MNRAS.420.1640R} stressed the effect of the amount of artificial viscosity added to resolve shocks, which can add an extra heating and prevent fragmentation. They also advised to use another implementation of the $\beta$-cooling called smoothed cooling.
\cite{2012MNRAS.427.2022M} have shown that changing the amount of artificial viscosity indeed changes the value of $\beta_{\text{crit}}$, and from runs with very high resolution (16 millions of SPH particles) they predict a converged value around 20. \cite{2014MNRAS.438.1593R} extended the work of \cite{2012MNRAS.420.1640R} and exhibited a converged $\beta_\mathrm{crit}$ between 6 and 8 with adjusted artificial viscosity parameters and smoothed cooling.

Other factors can change the determination of the fragmentation boundary. \cite{2007MNRAS.381.1543C} used the same set-up as in \cite{2005MNRAS.364L..56R} but reduced progressively the value of $\beta$ instead of setting it from the beginning. They found a value of $\beta_{\mathrm{crit}}$ two times lower than \cite{2005MNRAS.364L..56R}. 
\cite{2011MNRAS.416L..65P}  stress the influence of using smooth initial conditions, suggesting to use relaxed initial conditions instead. In a later work, \cite{2012MNRAS.421.3286P} have found than disc can fragment even for very high value of $\beta$ and argue that fragmentation is a stochastic process, with the probability of forming fragments diminishing for high value of $\beta$. \cite{2015MNRAS.451.3987Y} proposed a formula for the probability of forming surviving fragments by measuring the wait time between destroying shocks in 2D simulations.
More recently, \cite{2017A&A...606A..70K} suggested that oversteepening by the slope limiting function in 2D grid simulation may increase the value of $\beta_{\text{crit}}$ and also cause the stochastic fragmentation.
\cite{2017ApJ...847...43D} claim they obtained exact convergence using a Godunov solver on a meshless finite mass scheme, with $\beta_{\text{crit}} \gtrsim 3$.
Spiral or annular structures are found in all the global disc simulations mentioned above, as well in simulation using a constant value of $t_\mathrm{cool}$ throughout the disc \citep{2005ApJ...619.1098M, 2012ApJ...746...98M}.

This academic problem is of interest because it may help to understand the subtle interplay between the differential rotation, the self-gravity and the thermal processes in real discs. 
Even without considering a lot of the physics actually involved, the problem shows a deep complexity that has not been cleared out by the numerous previous studies in the literature. 

In this paper, we present 3D simulations of a disc undergoing Gammie's $\beta$-cooling using a Riemann solver and a Godunov scheme on a 3D grid.
By doing this, we get rid of the difficulty of setting manually the value of the artificial viscosity. Since the artificial viscosity seems to have a big influence on the fragmentation of the disc, it is important to study the convergence of the fragmentation boundary in this regime. 
We found that the distinction between fragmented and unfragmented discs is unclear and indeed does not seem to converge with resolution. 
If we also keep in mind \cite{2012MNRAS.421.3286P}'s claim that fragmentation is a stochastic process, it may be more relevant to study how do the properties of the discs evolve when we change $\beta$ rather than trying to draw a clear line between two supposedly separated regimes. 
That motivated us to study the column density fluctuation probability density function ($\Sigma$-PDF). It appears to show a power-law profile, with a slope that smoothly depends on $\beta$. The dependence on $\beta$ can be simply explained with the $\alpha$ disc formalism and can be used to build an empirical determination of the probability to fragment as a function of $\beta$.
We also found that the setting of the $\Sigma$-PDF and the of fragmentation are the result of a two-step process. 
First a filamentary spiral pattern (FSP) is created because of the gravitational instabilities and stabilized by the differential rotation. 
Secondly, the gas collapses alongside the filaments. A similar scenario was previously proposed by  \cite{2016MNRAS.458.3597T}. Our work brings a more precise description of the formation of the filament in the first step, and a probabilistic estimation of the amount of fragmentation expected in the second step.

The paper is organized as follows. In section \ref{sec:simu}, we present our numerical method and the set of simulation we ran. 
The global results for the fragmentation boundary and the $\Sigma$-PDF are presented in section \ref{sec:result}. We then present our two-step scenario with first the formation of filamentary structure (\ref{sec:stepone}) and then their collapse into fragments (\ref{sec:steptwo}). We draw our conclusions in section \ref{sec:conclusions}.
A summary of the main notations used is available in Table~\ref{tbl:notation}.

\begin{figure*}
\centering
\includegraphics[width=\textwidth]{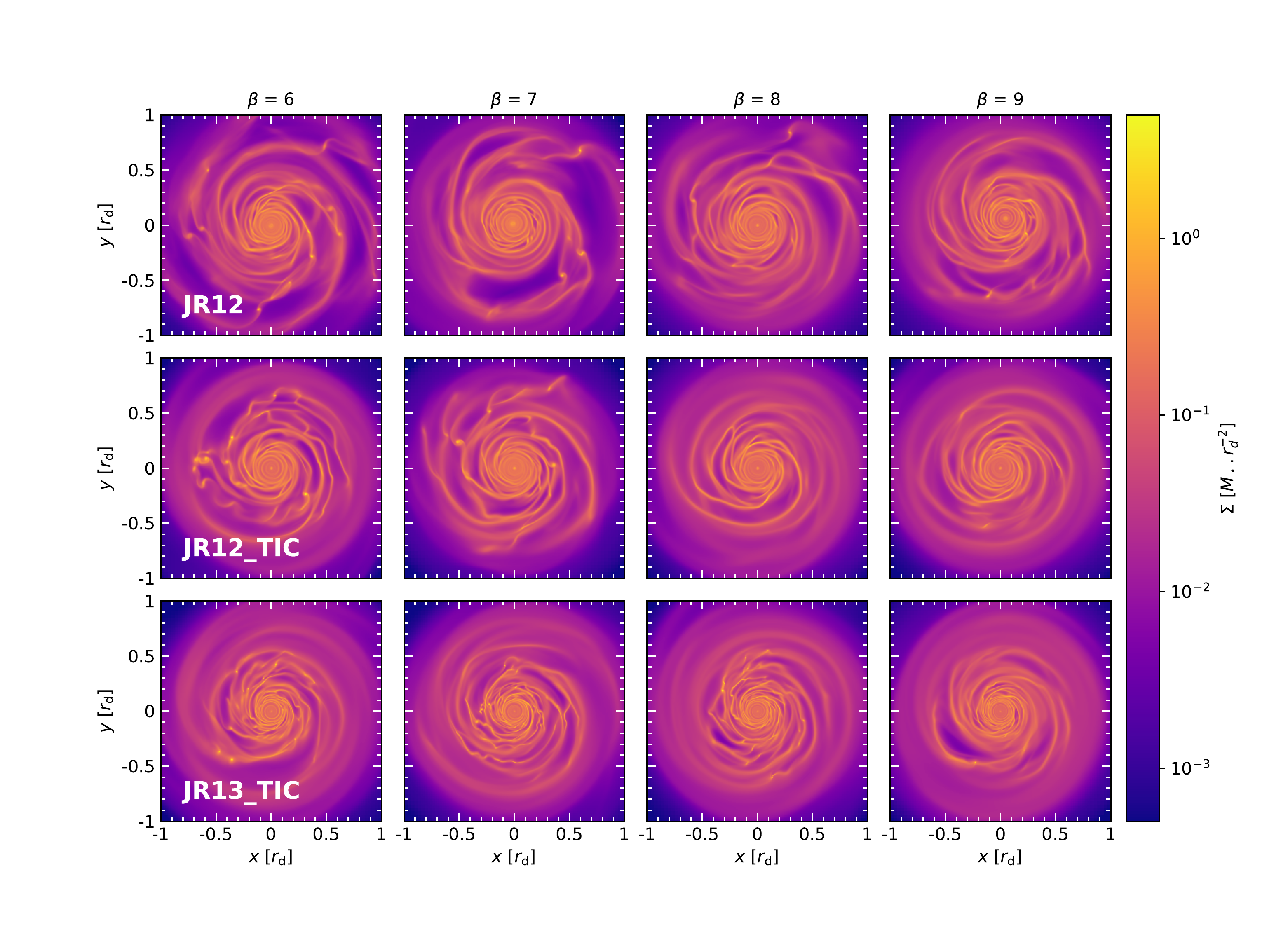} 

\caption{Column density map for different resolutions and value of $\beta$ at about $4.5$ ORP. 
A structure of spiral arms is visible. Fragments are visible respectively for $\beta \leq 7$ and $\beta \leq 9$ for the \textsc{JR12} and \textsc{JR13\_TIC}. The simulation \textsc{JR12\_TIC} at $\beta = 8$ fragments at a later time.}\label{fig:coldens}

\end{figure*}

\section{Simulation of self-gravitating  discs with $\beta$-cooling}
\label{sec:simu}
\subsection{Numerical set-up}
\label{subsec:set-up}

We simulate a disc of gas undergoing purely hydrodynamic forces, its own gravity and the $\beta$-cooling. The simulation is ran with the 3D-grid code \textsc{Ramses} \citep{2002A&A...385..337T} that uses a Godunov scheme. The flux between each cell is computed with the HLLC Riemann solver. The gravity potential is updated at each time-step with a Poisson solver, and a source term is added to the energy equation to implement the $\beta$-cooling.
The Euler equations write
\begin{align}
\label{eq:cont}
\frac{\partial \rho}{\partial t}+\bm{\nabla\cdot}(\rho \bm{u}) &=0,\\
\label{eq:euler_gen}
\frac{\partial \bm{v}}{\partial t} + \bm{u\cdot\nabla v} &= - \frac{1}{\rho}\bm{\nabla}P + \bm{g_\star} + \bm{g_{\mathrm{gas}}},\\
\label{eq:cooling}
\dfrac{\partial E}{\partial t} + \nabla \left( \left( E + P \right) 
\bm{v} \right) &= - \dfrac{U}{t_{\text{cool}}},
\end{align}
where $E$ is the total energy, $P$ is the pressure, $\bm{v}$ is the velocity,  $U$ is the internal energy, and $ \bm{g_\star}$ and $\bm{g_{\mathrm{gas}}}$ are the gravitational fields of the star and the gas, respectively.

\subsection{Initial and boundary conditions}

We use the same initial conditions as in \cite{2012MNRAS.427.2022M} to allow comparison. 
The specific disc set-up for \textsc{Ramses} was inspired by \cite{2017A&A...599A..86H}.
The disc is initially close to equilibrium with an initial column density profile $\Sigma \propto r^{-1}$ and a temperature profile $T \propto r^{-1/2}$ where $r$ is the cylindrical radius. The disc has a radius $r_d = 0.25$ (code units), after which the density is divided by 100. The density and temperature at the disc radius $r_d$ are chosen so that the mass of the disc is $M_d = 0.1 M_\star$, where $M_\star$ is the mass of the central object, and the initial value of the Toomre parameter at the disc radius is $Q_{0,d} = 2$. The adiabatic index of the gas is $\gamma = 5/3$.

The simulation is run within a cube of size $L=2$. Although the problem has a cylindrical symmetry, we use Cartesian coordinates. This prevents having a singularity at the centre of the box but has several caveats. 
The first concern with the Cartesian grid is the poor conservation of the total angular momentum \citep{2015A&A...579A..32L}. \cite{2017A&A...599A..86H} investigated this issue in a set-up similar to ours with a slightly lower resolution and found that the value of $\alpha$ induced by the loss or gain of angular momentum is under $10^{-3}$  (see their Figure 2), which is well below the typical value of fragmenting discs (a few $10^{-2}$ according to \cite{2005MNRAS.364L..56R}). 
Similar measurement were made with our specific set-up but with a gravitationally stable disc ($Q > 3$), in appendix~\ref{sec:consangmom}. We found an even lower value for the numerical $\alpha$, below $10^{-4}$ (Figure~\ref{fig:alphaQ3}).
Another caveat is the poor resolution on the centre of the cube but this is mitigated by the use of the adaptive mesh refinement (AMR).
Finally, having a cubic box may introduce spurious reflection at the border of the simulation.
To avoid this, we maintain a dead zone over a radius of $0.875$ (in code units) where all variables are replaced by their initial value at each time-step. This method has been used in \cite{2017A&A...599A..86H} and has proven to be efficient.

\subsection{Simulations}
\label{subsec:simu}

\begin{table}
\caption{List of simulations. The fragmented column corresponds to Definition 1 (see the text). }\label{tbl:simu}
  \begin{center}
	\begin{tabular}{clccc}
	\hline\noalign{\smallskip}
	Group & Name & $l_{max}$ & $\beta$ & Fragmented \\
	\noalign{\smallskip}
	\hline
	\noalign{\smallskip}
	\multirow{5}{*}{\textsc{JR11}} 
	& \texttt{beta4\_jr11} & 11 & $4$ & \textbf{Yes} \\
	&\texttt{beta5\_jr11} & 11 & $5$ & \textbf{Yes} \\
	&\texttt{beta6\_jr11} & 11 & $6$ & No \\
	&\texttt{beta7\_jr11} & 11 & $7$ & No \\
	&\texttt{beta8\_jr11} & 11 & $8$ & No \\
	\hline
	\multirow{14}{*}{\textsc{JR12}} 
	&\texttt{beta2\_jr12} & 12 & $2$ & \textbf{Yes} \\
	&\texttt{beta3\_jr12} & 12 & $3$ & \textbf{Yes} \\
	&\texttt{beta4\_jr12} & 12 & $4$ & \textbf{Yes} \\
	&\texttt{beta5\_jr12} & 12 & $5$ & \textbf{Yes} \\
	&\texttt{beta6\_jr12} & 12 & $6$ & \textbf{Yes}  \\
	&\texttt{beta7\_jr12} & 12 & $7$ & \textbf{Yes} \\
	&\texttt{beta8\_jr12} & 12 & $8$ & \textbf{Yes}  \\
	&\texttt{beta9\_jr12} & 12 & $9$ & No  \\
	&\texttt{beta10\_jr12} & 12 & $10$ & No \\
	&\texttt{beta11\_jr12} & 12 & $11$ & No \\
	&\texttt{beta12\_jr12} & 12 & $12$ & No \\
	&\texttt{beta14\_jr12} & 12 & $14$ & No \\
	&\texttt{beta16\_jr12} & 12 & $16$ & No \\
	&\texttt{beta18\_jr12} & 12 & $18$ & No \\
	\hline
	\multirow{3}{*}{\textsc{JR12\_TIC}} 
	&\texttt{beta6\_jr12\_tic} & 12 & $6$ & \textbf{Yes} \\
	&\texttt{beta8\_jr12\_tic} & 12 & $8$ & \textbf{Yes}  \\
	&\texttt{beta9\_jr12\_tic} & 12 & $9$ & No \\
	&\texttt{beta10\_jr12\_tic} & 12 & $10$ & No \\
	&\texttt{beta12\_jr12\_tic} & 12 & $12$ & No \\
	\hline
	\multirow{6}{*}{\textsc{JR13\_TIC}} 
	&\texttt{beta4\_jr13\_tic} & 13 & $4$ & \textbf{Yes} \\
	&\texttt{beta6\_jr13\_tic} & 13 & $6$ & \textbf{Yes} \\
    &\texttt{beta7\_jr13\_tic} & 13 & $7$ & \textbf{Yes} \\
    &\texttt{beta8\_jr13\_tic} & 13 & $8$ & \textbf{Yes} \\
    &\texttt{beta9\_jr13\_tic} & 13 & $9$ & \textbf{Yes} \\
    &\texttt{beta10\_jr13\_tic} & 13 & $10$ & No \\
    &\texttt{beta12\_jr13\_tic} & 13 & $12$ & No \\
    &\texttt{beta14\_jr13\_tic} & 13 & $14$ & No \\
    &\texttt{beta16\_jr13\_tic} & 13 & $16$ & No \\ 

	\noalign{\smallskip}
	\hline
	\end{tabular} 
  \end{center}
\end{table}

One of the goals of this work is to study the convergence of the fragmentation boundary with a Godunov scheme. 
To do so, we run simulations for several values of $\beta$ and several resolutions. To reduce the computation time, we use the \textsc{Ramses}'s Adaptative Mesh Refinement (AMR). The level of refinement of a cell is the number of times the simulation box must be divided in eight equal part to get the cell. 
The minimal level of refinement is 8, meaning that the root grid has a size of $256^3$. The disc itself is refined at a minimum level of $10$.
Only the parts of the simulation which are prone to form fragments are simulated with full resolution. 
Each cell is refined until the Jeans's length is covered by at least 20 cells or it reaches the maximum level of refinement $l_{\max}$. 
Thus, the resolution of a simulation is given by the value of $l_{\max}$.
Table~\ref{tbl:simu} lists the simulations that were run for this study.
A first set of simulations with $l_{\max} = 11$ to $l_{\max} = 12$ are run until about 5 Outer Rotation Periods (ORP); that is, that the gas at the border of the disc had 5 orbits around the star.
A second set of simulations, labelled \texttt{tic}, for Turbulent Initial Condition, were run from relaxed initial conditions for $l_{\max} = 12$ and $l_{\max} = 13$.
More precisely, they were restarted from a simulation at $\beta = 20$ and $l_{\max} = 12$ for which the whole disc reached a gravitoturbulent state (after two ORPs).
According to \cite{2011MNRAS.416L..65P} and \cite{2007MNRAS.381.1543C}, departing from such turbulent conditions should reduce spurious fragmentation. 


\section{Results}
\label{sec:result}

Figure~\ref{fig:coldens} features column density maps for $\beta$ between $6$ and $9$ and a resolution $l_{\max}$ from $12$ to $13$. 
For each of these simulations, a structure of spiral arms develops. 
As we shall see it what follows, the filamentary nature of these arms is important. We will thus refer to it as the Filamentary Spiral Pattern (FSP). When the disc is fragmented, the bounded fragments appear within the FSP. 
They are more frequent for low value of $\beta$ and high value of the resolution. 
Gas nearby to these clumps are locally orbiting around them.

\subsection{The fragmentation boundary}
\label{subsec:boundary}

In this section we determine a fragmentation boundary to compare with previous result in the literature.
There is no clear separation between the fragmented and unfragmented cases, and we shall see that the boundary depend on the definition chosen for a fragment as well as on the resolution.

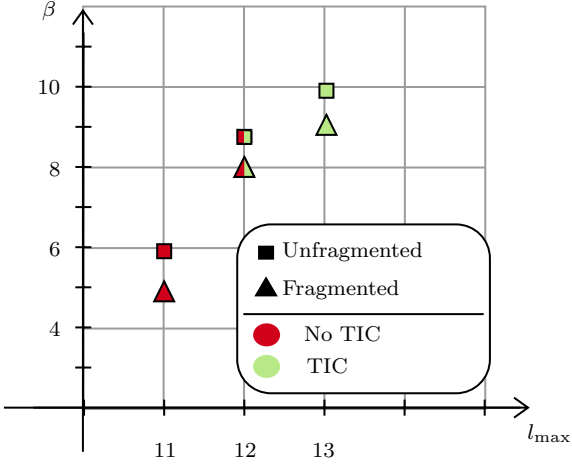
\begin{figure}
	\begin{tikzpicture}[x=0.75pt,y=0.75pt,yscale=-1,xscale=1]

\draw  [draw opacity=0] (120.57,25) -- (321.28,25) -- (321.28,225.71) -- (120.57,225.71) -- cycle ; \draw  [color={rgb, 255:red, 155; green, 155; blue, 155 }  ,draw opacity=1 ] (160.71,25) -- (160.71,225.71)(200.85,25) -- (200.85,225.71)(240.99,25) -- (240.99,225.71)(281.14,25) -- (281.14,225.71) ; \draw  [color={rgb, 255:red, 155; green, 155; blue, 155 }  ,draw opacity=1 ] (120.57,65.14) -- (321.28,65.14)(120.57,105.28) -- (321.28,105.28)(120.57,145.43) -- (321.28,145.43)(120.57,185.57) -- (321.28,185.57) ; \draw  [color={rgb, 255:red, 155; green, 155; blue, 155 }  ,draw opacity=1 ] (120.57,25) -- (321.28,25) -- (321.28,225.71) -- (120.57,225.71) -- cycle ;
\draw  [fill={rgb, 255:red, 255; green, 255; blue, 255 }  ,fill opacity=1 ] (197.46,150.47) .. controls (197.46,141.17) and (205,133.64) .. (214.3,133.64) -- (306.75,133.64) .. controls (316.05,133.64) and (323.59,141.17) .. (323.59,150.47) -- (323.59,200.99) .. controls (323.59,210.29) and (316.05,217.83) .. (306.75,217.83) -- (214.3,217.83) .. controls (205,217.83) and (197.46,210.29) .. (197.46,200.99) -- cycle ;
\draw  [fill={rgb, 255:red, 0; green, 0; blue, 0 }  ,fill opacity=1 ] (208.36,144.74) -- (215.36,144.74) -- (215.36,150.71) -- (208.36,150.71) -- cycle ;
\draw  [fill={rgb, 255:red, 0; green, 0; blue, 0 }  ,fill opacity=1 ] (211.86,160.51) -- (216.86,169.04) -- (206.86,169.04) -- cycle ;
\draw    (200.52,178.37) -- (320.52,178.37) ;
\draw  [draw opacity=0][fill={rgb, 255:red, 208; green, 2; blue, 27 }  ,fill opacity=1 ] (205.41,188.73) .. controls (205.41,185.56) and (208.42,182.99) .. (212.13,182.99) .. controls (215.85,182.99) and (218.86,185.56) .. (218.86,188.73) .. controls (218.86,191.89) and (215.85,194.46) .. (212.13,194.46) .. controls (208.42,194.46) and (205.41,191.89) .. (205.41,188.73) -- cycle ;
\draw  [draw opacity=0][fill={rgb, 255:red, 184; green, 233; blue, 134 }  ,fill opacity=1 ] (205.41,204.45) .. controls (205.41,201.29) and (208.42,198.72) .. (212.13,198.72) .. controls (215.85,198.72) and (218.86,201.29) .. (218.86,204.45) .. controls (218.86,207.62) and (215.85,210.19) .. (212.13,210.19) .. controls (208.42,210.19) and (205.41,207.62) .. (205.41,204.45) -- cycle ;
\draw  (95.75,225) -- (342.63,225)(120.44,27) -- (120.44,247) (335.63,220) -- (342.63,225) -- (335.63,230) (115.44,34) -- (120.44,27) -- (125.44,34)  ;
\draw    (81,225) -- (338.75,225) (121,221) -- (121,229)(161,221) -- (161,229)(201,221) -- (201,229)(241,221) -- (241,229)(281,221) -- (281,229)(321,221) -- (321,229) ;
\draw    (120.57,225) -- (120.75,38) (116.59,205) -- (124.59,205)(116.61,185) -- (124.61,185)(116.63,165) -- (124.63,165)(116.65,145) -- (124.65,145)(116.66,125) -- (124.66,125)(116.68,105) -- (124.68,105)(116.7,85) -- (124.7,85)(116.72,65) -- (124.72,65)(116.74,45) -- (124.74,45) ;
\draw  [fill={rgb, 255:red, 208; green, 2; blue, 27 }  ,fill opacity=1 ] (197.5,86.5) -- (204.5,86.5) -- (204.5,93.5) -- (197.5,93.5) -- cycle ;
\draw  [draw opacity=0][fill={rgb, 255:red, 184; green, 233; blue, 134 }  ,fill opacity=1 ] (201,86.78) -- (204.5,86.78) -- (204.5,93.06) -- (201,93.06) -- cycle ;
\draw    (197.5,93.5) -- (204.5,93.5) ;
\draw    (204.5,93.06) -- (204.5,86.78) ;

\draw  [fill={rgb, 255:red, 208; green, 2; blue, 27 }  ,fill opacity=1 ] (201,100) -- (206,110) -- (196,110) -- cycle ;
\draw  [draw opacity=0][fill={rgb, 255:red, 184; green, 233; blue, 134 }  ,fill opacity=1 ][line width=0.75]  (201,100) -- (206,109.92) -- (201,109.92) -- cycle ;
\draw    (201,100) -- (206,110) ;
\draw    (196,110) -- (206,109.92) ;

\draw  [fill={rgb, 255:red, 208; green, 2; blue, 27 }  ,fill opacity=1 ] (157.5,143.5) -- (164.5,143.5) -- (164.5,150.5) -- (157.5,150.5) -- cycle ;
\draw  [fill={rgb, 255:red, 208; green, 2; blue, 27 }  ,fill opacity=1 ] (161,162) -- (166,172) -- (156,172) -- cycle ;
\draw  [fill={rgb, 255:red, 184; green, 233; blue, 134 }  ,fill opacity=1 ] (238.5,63.5) -- (245.5,63.5) -- (245.5,70.5) -- (238.5,70.5) -- cycle ;
\draw  [fill={rgb, 255:red, 184; green, 233; blue, 134 }  ,fill opacity=1 ] (242,79) -- (247,89) -- (237,89) -- cycle ;
\draw    (197.5,86.5) -- (204.5,86.78) ;

\draw (353.36,237.6) node   [align=left] {$\displaystyle l_{\max}$};
\draw (103.5,27) node   [align=left] {$\displaystyle \beta $};
\draw (201.5,245.96) node   [align=left] {$\displaystyle 12$};
\draw (106.5,186) node   [align=left] {$\displaystyle 4$};
\draw (106.5,146) node   [align=left] {$\displaystyle 6$};
\draw (106.5,106) node   [align=left] {$\displaystyle 8$};
\draw (240.5,245.96) node   [align=left] {$\displaystyle 13$};
\draw (161.5,245.96) node   [align=left] {$\displaystyle 11$};
\draw (103.5,65) node   [align=left] {$\displaystyle 10$};
\draw (255,147) node   [align=left] {Unfragmented};
\draw (255,166) node   [align=left] {Fragmented~~~};
\draw (250,188) node   [align=left] {No TIC};
\draw (250,204) node   [align=left] {TIC~~~~};

\end{tikzpicture}
    \caption{Fragmentation limit as a function of the resolution}
    \label{fig:fragbound_d1}
\end{figure}

\paragraph*{Definition 1:} We define a fragment as a zone where the column density $\Sigma$ is 30 times higher than the mean azimuthal column density $\overline{\Sigma}$, which survives for at least one orbital period.
This definition was inspired by \cite{2012MNRAS.421.3286P}. Note that changing the threshold of 30 or the minimum survival time may change the value of the fragmentation boundary, and thus this definition is quite unsatisfactory.
However, this should not influence the convergence result. The fragmentation result is reported on the table \ref{tbl:simu}, allowing us to get an upper and lower estimate of the fragmentation boundaries at a given resolution. These estimates are presented in Figure~\ref{fig:fragbound_d1}. We find a fragmentation boundary which is near $\beta = 9$ for our highest resolution. This is close to values previously found in the literature for 3D simulations with SPH \citep{2011MNRAS.410..559M, 2012MNRAS.427.2022M}. 
At coarser resolution the fragmentation boundary lies at a lower value of $\beta$.
There is no change of the boundary when TIC are used for the intermediate resolution.

\paragraph*{Definition 2:} As discussed above, \emph{Definition 1} is unsatisfactory because the results depend on free parameters. We introduce another definition inspired by
\cite{2003MNRAS.339.1025R} and compare the results. The definition is the following: a fragment is a gravitationally bound clump of gas, that is for which the sum of the internal energy and its self-gravitational energy is negative and which survives for at least one orbital period.
We used the HOP algorithm \citep{1998ApJ...498..137E} to find such clumps and select the ones that fulfill the energy criterion. Clumps are then followed from one snapshot to the other to determine if it survives more than one orbital period. We tested this method for the group \textsc{JR12} a found bounded clump for the simulation with $\beta = 9$, the simulation with $\beta = 10$ being still unfragmented according to this definition.
The determined fragmentation boundary for this resolution is by consequence slightly superior when Definition 2 is used.

\paragraph*{} We conclude that the simulations of a $\beta$-cooling with a Riemann solver and a Godunov scheme we ran yield fragmentation boundary result close from what was found with other techniques in the literature.
The fragmentation boundary seems to lie between $\beta = 8$ and $\beta = 10$, and is not well converged with resolution. The two definitions we used for a fragment, borrowed from older publications, led to two different (but close) determinations of the fragmentation boundary.

\subsection{Column density probability density function}
\label{subsec:pdf}

\begin{figure}
\begin{center}
\includegraphics[width=0.5 \textwidth]{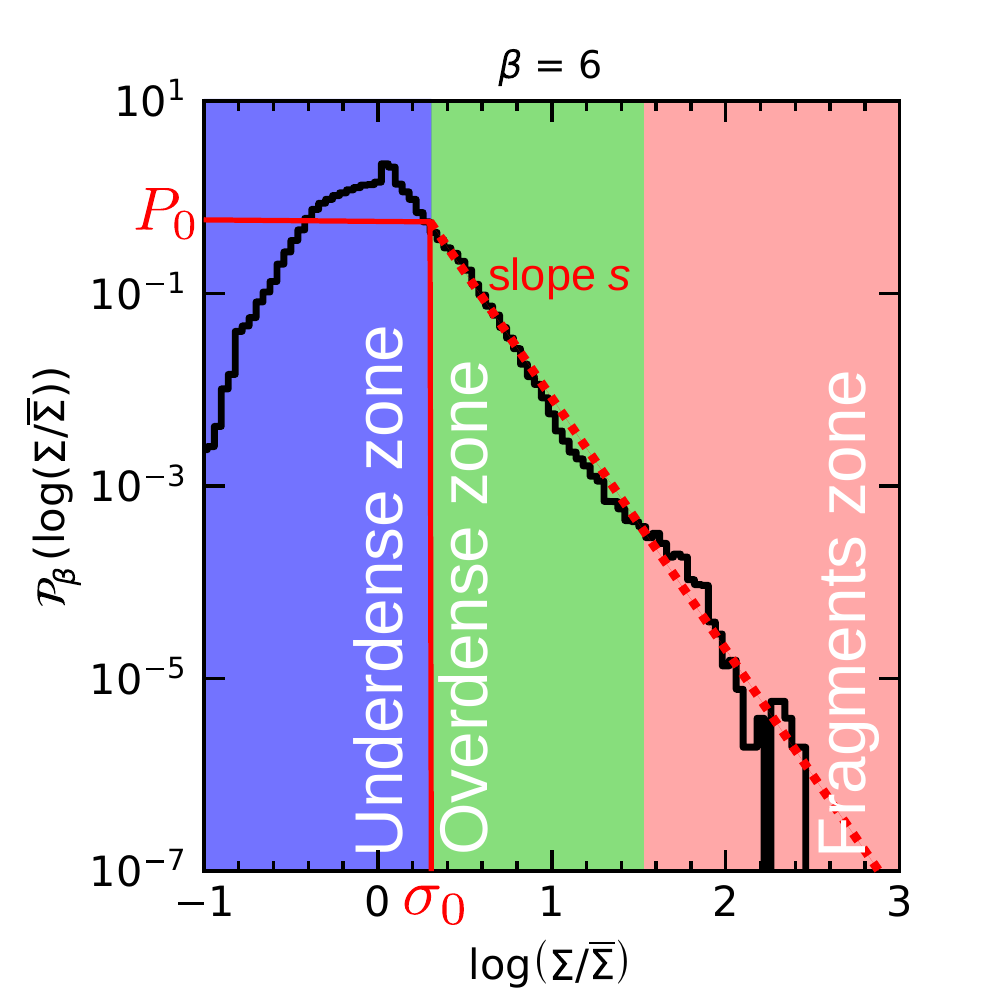} 
\caption{The time-averaged $\Sigma$-PDF (see the text) for the simulation $\beta = 6$ of the group \textsc{JR12}. The \emph{overdense zone} is well fitted by a power-law.}
\label{fig:pdf_expl}
\end{center}
\end{figure}

Both the results above and the long-lasting debate in the literature about the value and the convergence of the fragmentation boundary make it hard to consider the fragmentation boundary as a reliable criterion for the fragmentation in discs. It is well-established that increasing the time-scale at which the disc is cooling (by increasing $\beta$) makes it less likely to fragment. However, it is unclear that there is a well-defined limit between the fragmented and the unfragmented regimes and the transition seems to be rather smooth.
As a consequence, a more statistical approach should be used to study this transition.

The density fluctuation probability function ($\rho$-PDF) is a powerful tool widely used in the context of the fragmentation of the interstellar medium  \citep{2012A&ARv..20...55H}. 
In the context of astrophysical discs, $\rho$-PDF has been used by \cite{2013MNRAS.430.1653H} to argue that the fragmentation can occur for any value of $\beta$ in theory.
However, the $\rho$-PDF of a simulated disc is hard to interpret because of the disc's strong radial and vertical structure. To get rid of the intrinsic fluctuations due to the structure, we use the probability function of the logarithm of the column density fluctuations $\sigma = \Sigma/\overline{\Sigma}$ with respect to its azimuthal average. In this paper, we call it $\Sigma$-PDF and note it $\mathcal{P}_\beta(\log(\sigma))$.
Using the column density suppresses the fluctuations due to the vertical stratification and averaging azimuthally instead of globally removes the effects of the radial structure.

\begin{figure}
\centering
\includegraphics[width=0.5 \textwidth]{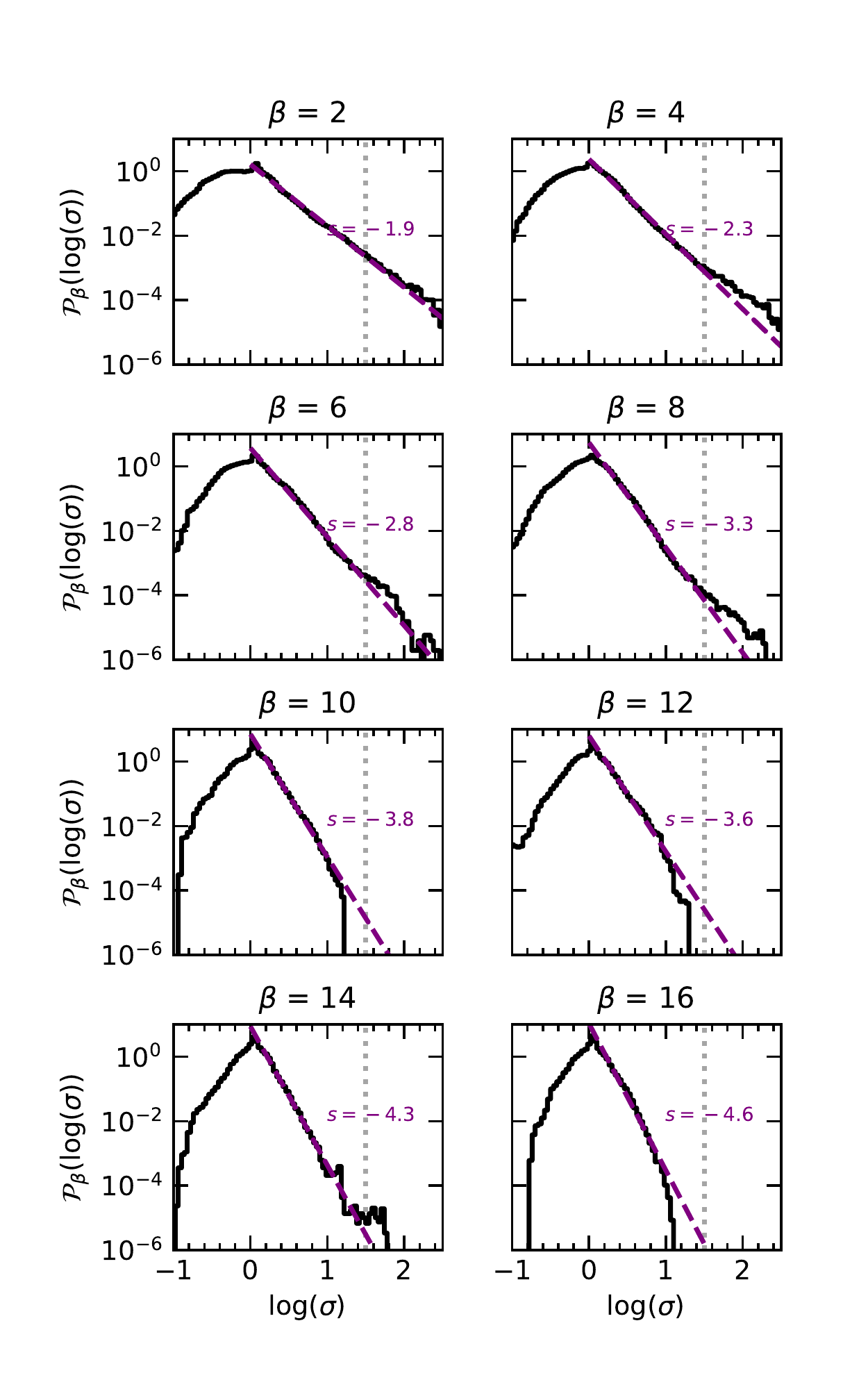} 

\caption{Time-averaged $\Sigma$-PDF for several different values of $\beta$ (group \textsc{JR12}) with $\sigma = \Sigma / \overline{\Sigma}$. The value of the PDF is averaged bin by bin. The dashed purple line is a fit of the overdense zone (see Figure~\ref{fig:pdf_expl}) and the vertical dotted grey line corresponds to the limit between the overdense zone and the fragments zone, in accordance with Definition 1.}\label{fig:pdf}

\end{figure}

Figure~\ref{fig:pdf_expl} shows an example of such a $\Sigma$-PDF. The graph can be separated in three regions. First the \emph{average and underdense zone} ($\sigma < \sigma_0$) presents approximately a log-normal shape,  then the \emph{overdense zone} ($\sigma_0 < \sigma <  \sigma_\mathrm{frag}$) has a power-law shape and finally a third one  ($\sigma >  \sigma_\mathrm{frag}$) may correspond to the fragments. 
The value of $\sigma_0$ is also chosen so that the value $P_0 = \mathcal{P}_\beta(\log(\sigma_0))$ is the same for all the simulations independently of $\beta$ (Figure~\ref{fig:sigma0_P0}). We will see in the section \ref{subsec:sbeta_model} that this value may actually be of prime importance. The value of $\sigma_\mathrm{frag}$ is 30 in accordance with Definition~1. Figure~\ref{fig:pdf} features the time-averaged $\Sigma$-PDF for the group \textsc{JR12}. 

Power-laws in $\rho$-PDF of molecular clouds are the signature of a dominating self-gravity  but in our particular situation the differential rotation may also play a role.

\subsection{Relationship between the slope of the Σ-PDF and β}
\label{subsec:sbeta}

\begin{figure}
\begin{center}
\includegraphics[width=0.5 \textwidth]{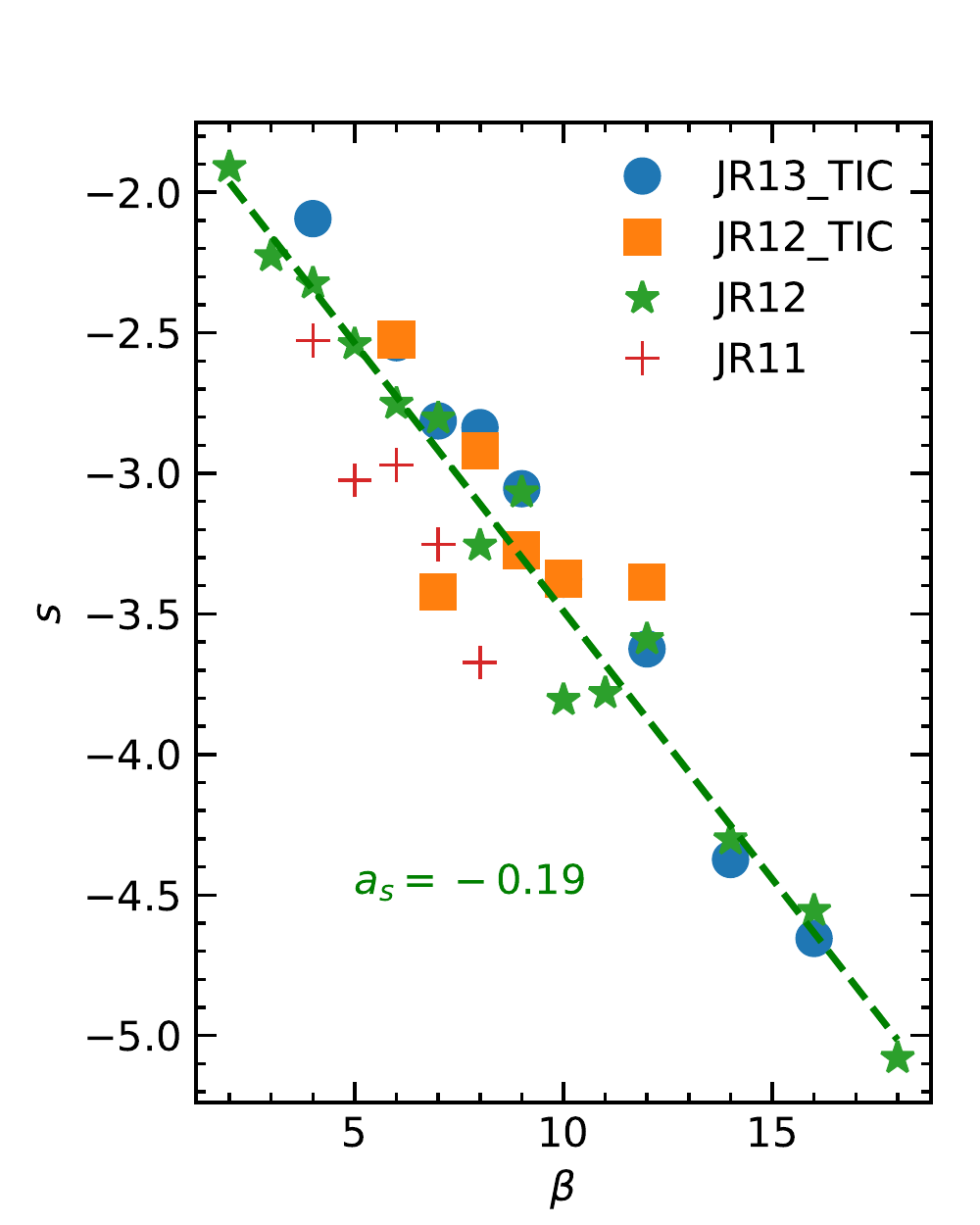} 
\caption{Slope $s$ of the overdense region of the $\Sigma$-PDF (see Figures \ref{fig:pdf_expl} and \ref{fig:pdf}) as a function of $\beta$ (same notations as in Equation~\eqref{eq:pwlw}. The dashed line is a linear fit of the data from the group \textsc{JR12} (Equation~\eqref{eq:linship}).}
\label{fig:kappabeta}
\end{center}
\end{figure}

\begin{figure}
\begin{center}
\includegraphics[width=0.5 \textwidth]{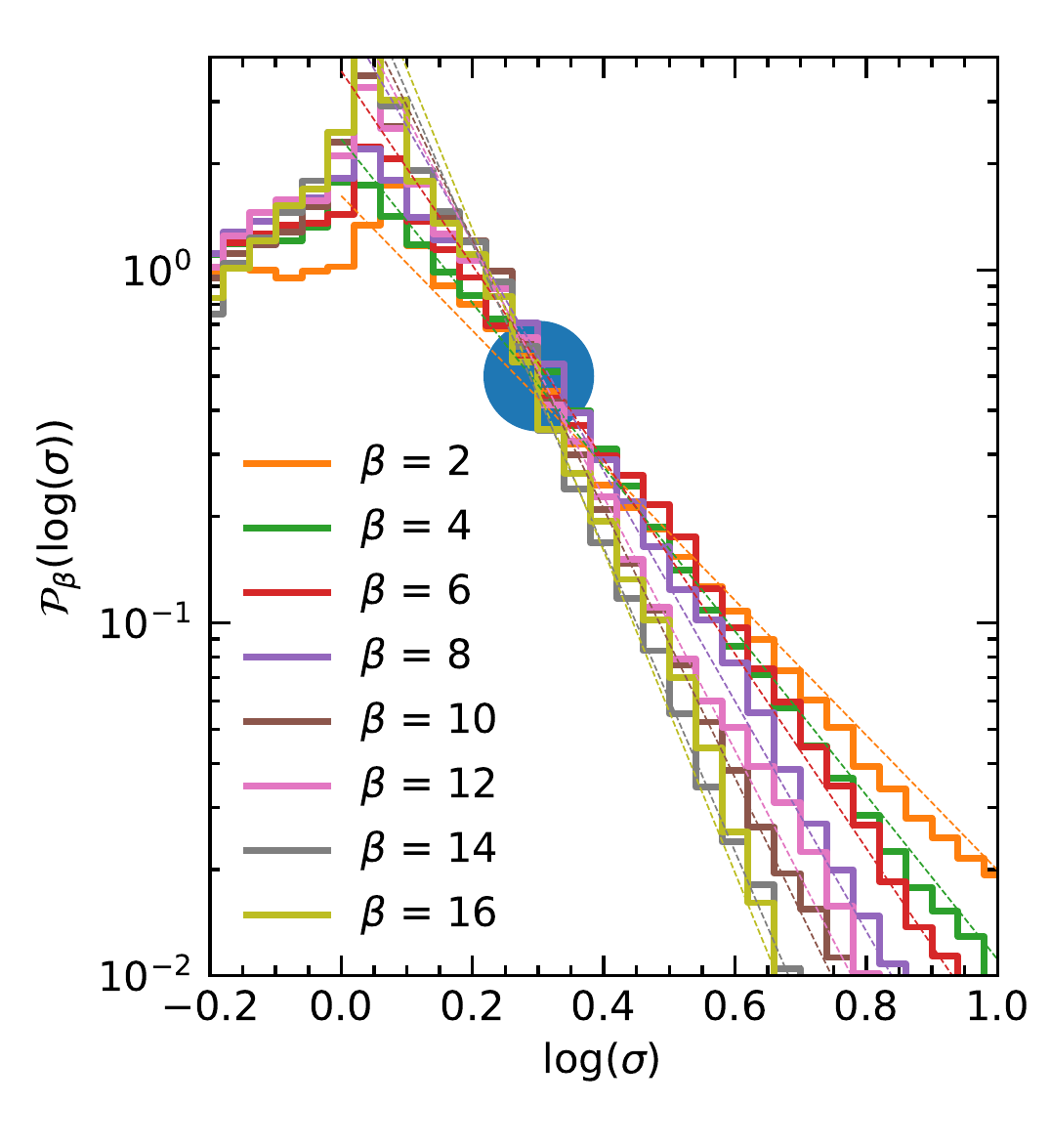} 
\caption{Zoom over the PDFs of Figure~\ref{fig:pdf}. The value of the PDF is roughly the same for $\sigma_0 = 2$ and is $P_0 \approx 0.5$ (blue area).}
\label{fig:sigma0_P0}
\end{center}
\end{figure}

\begin{figure*}
\begin{center}
\includegraphics[width=0.45 \textwidth]{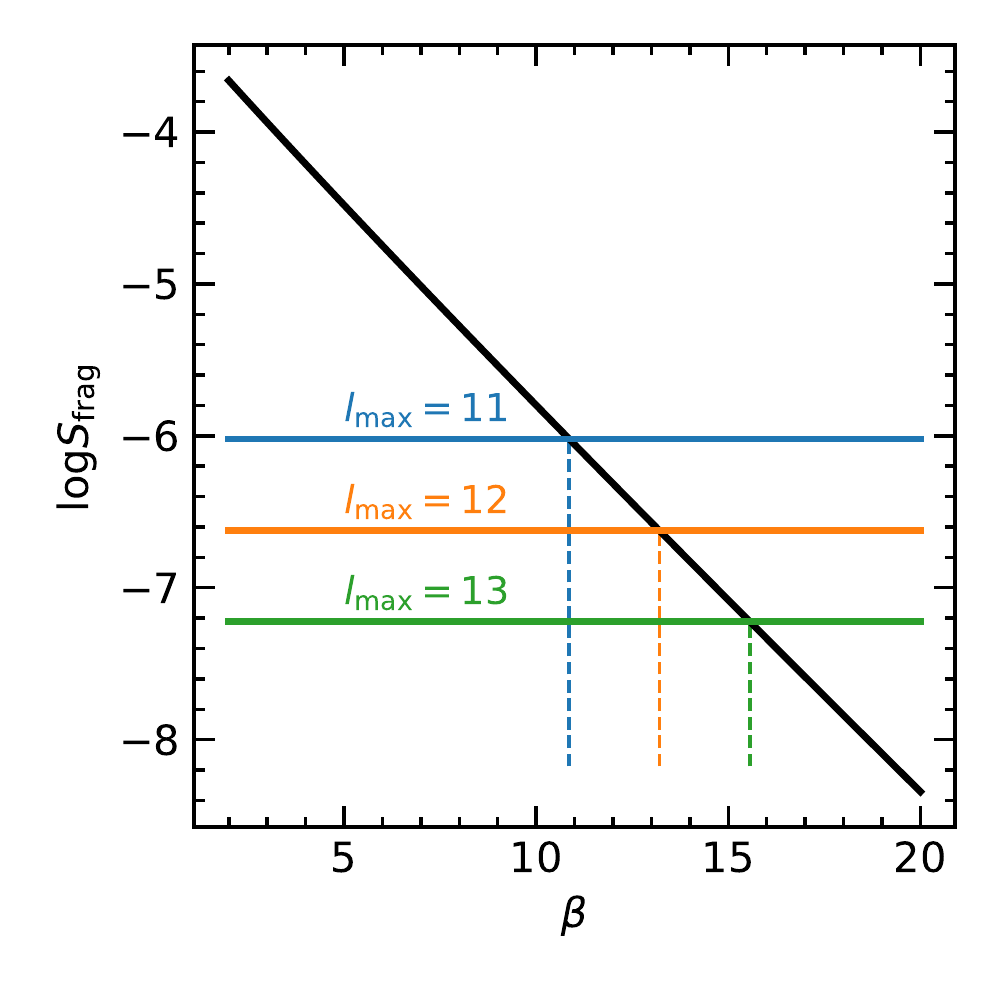} 
\includegraphics[width=0.45 \textwidth]{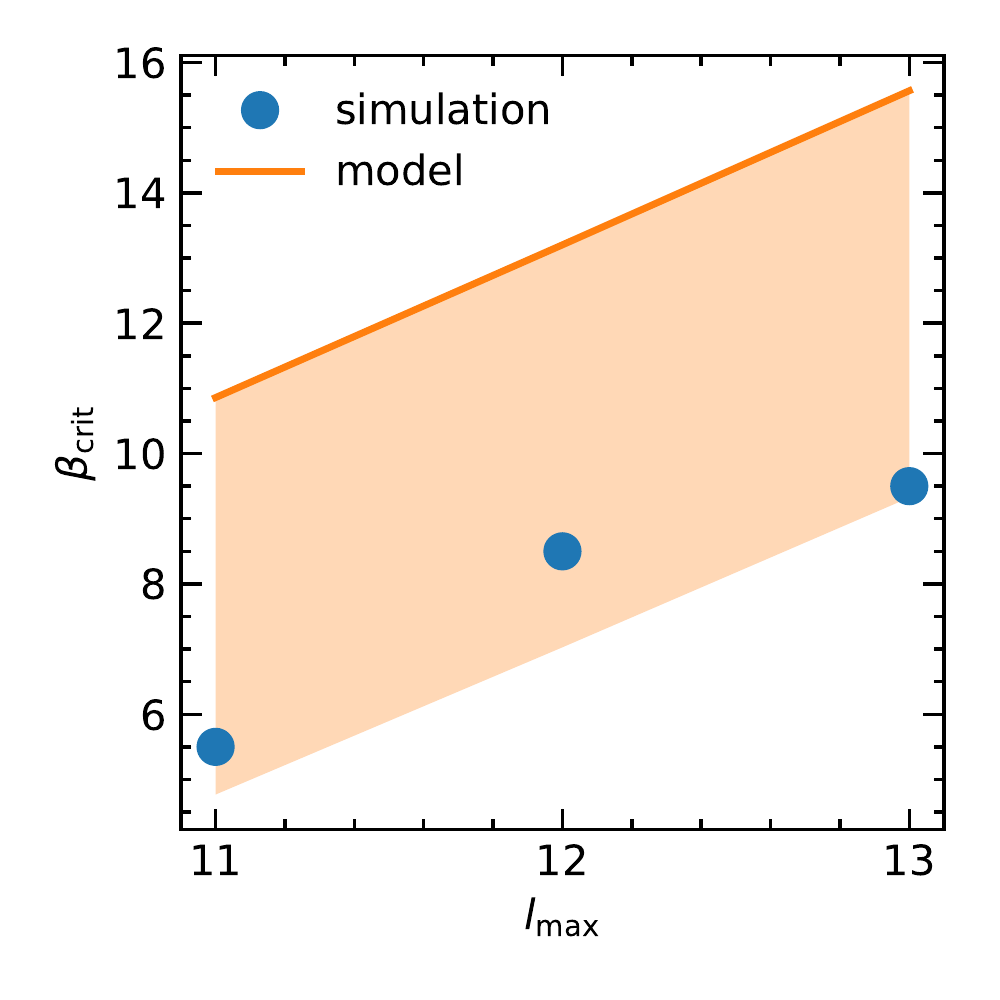} 

\caption{\textbf{Left}: Surface $S_\mathrm{frag}$ of the disc with a column density over $\sigma_\mathrm{frag} = 30$ as a function of $\beta$ according to Equation~\eqref{eq:pfrag} (in black). The coloured lines correspond to the tinniest resolved surface at the indicated level. The vertical dotted lines give an upper bound for $\beta_\mathrm{crit}$ at the given resolution, that is a value of $\beta$ for which the fragmented zone (as defined in Figure~\ref{fig:pdf_expl}) is resolved by at least one cell. \textbf{Right}: Comparison between the upper bound for $\beta_\mathrm{crit}$ from our empirical model (left-hand panel) and the values measured in the simulations. The bottom of the shaded zone corresponds to the maximal value of $\beta$ for which the fragmented zone is resolved by at least 40 cells according to the model. }
\label{fig:sfrag}
\end{center}
\end{figure*}

From Figure~\ref{fig:pdf}, we can see that the slope of the $\Sigma$-PDF strongly depends on the cooling parameter $\beta$. To measure this, we define $s$ the slope of the overdense zone of the time-averaged $\Sigma$-PDF ($\sigma_0 < \sigma < \sigma_\mathrm{frag}$):
\begin{equation}
\label{eq:pwlw}
\mathcal{P}_\beta\left(\log\left(\sigma\right)\right) = P_0 \left(\dfrac{\sigma}{\sigma_0}\right)^{s}.
\end{equation}
We compute $s$  from the simulation by fitting a power-law on the $\Sigma$-PDF.
Slow cooling (that is high value of $\beta$) led to higher value of $s$ as highlighted in Figure~\ref{fig:kappabeta}. There seems to be a linear relationship between $s$ and $\beta$. 
A linear fit of the data from the group \textsc{JR12} yields the following relationship.
\begin{align}
\label{eq:linship}
  &  s = a_s \beta + b_s, & a_s \approx -0.2,  b_s \approx -1.6,  R^2 = 0.97,
\end{align}
where  $R$, the multiple correlation coefficient of the fit, is remarkably close to 1. The group \textsc{JR12} is chosen for the fit because it is the group for which we explored the widest range of value of $\beta$. 
As said above, $P_0 = \mathcal{P}_\beta(\log(\sigma_0))$ is approximately the same whatever the value of $\beta$ and approximately equal to $0.5$ (see Figure~\ref{fig:sigma0_P0}).

\subsection{Energy balance in the disc}
\label{subsec:sbeta_model}

In this section, we try to understand and interpret the dependence on $\beta$ of the slope of the $\Sigma$-PDF shown in Figure~\ref{fig:kappabeta}. We consider an annular ring between radius $r$ and $r + \Delta r$ (with $\Delta r \ll r$), of mean column density $\overline{\Sigma}$ and volume $V = 4 \pi h r \Delta r $ where $h$ is the scale height of the disc. 
In a self-gravitating disc, the main source of angular momentum transport is the self-gravitating torque (e.g., \citeauthor{2004MNRAS.351..630L}, \citeyear{2004MNRAS.351..630L}),
which can be expressed as 
\begin{equation}
\alpha_{\mathrm{grav}} = \dfrac{2}{3}\dfrac{1}{4 \pi G} \dfrac{2h}{\overline{\Sigma} c_s^2 V} \int_V g_r g_\varphi  \dif V.
\label{eq:alpha_grav}
\end{equation}
Since $\bm{\nabla}\cdot \bm{g} = - 4 \pi G \rho$ we approximate $g_\varphi$ to be equal to
\begin{equation}
g_\varphi \simeq  \varepsilon_\varphi  2 \pi G \Sigma.
\label{eq:approx_gphi}
\end{equation}
with $\varepsilon_\varphi$ an efficiency parameter which depends on the azimuthal anisotropies. 
For the radial component $g_r$, we assume that it is dominated by the gas gravity rather than by the stellar one. This is correct if the disc column
density is about 10–25 times larger than the mean disc column density (based on $r/h \simeq 3$–$5$). In the same fashion, we get
\begin{equation}
g_r \simeq \varepsilon_r 2 \pi G \Sigma.
\label{eq:approx_gr}
\end{equation}

As shown in Figure~\ref{fig:pdf_expl}, the column density PDF can be decomposed in a low column density part, which contains most of the mass and is not strongly affected by $\beta$ 
and a high column density one which is clearly a power-law and whose slope depends on $\beta$. Using Equation~\eqref{eq:pwlw}, the PDF for $\Sigma$
when $\Sigma > \sigma_0 \overline{\Sigma}$ can be written as
\begin{equation}
\mathrm{PDF}(\Sigma) = \dfrac{P_0}{\sigma_0 \overline{\Sigma}}  \left(\dfrac{\Sigma}{\sigma_0 \overline{\Sigma}}\right)^{s-1}.
\end{equation}
We further assume that the dominant contribution to $\alpha_{\mathrm{grav}}$ comes from 
this part of the PDF. We can rewrite Equation~\eqref{eq:alpha_grav} by summing over $\Sigma$ instead of the volume that leads to
\begin{equation}
\alpha_{\mathrm{grav}} 
= \dfrac{2}{3} \dfrac{h}{2 \pi G c_s^2 } \int _{\sigma_0 \overline{\Sigma}} ^\infty \dfrac{P_0}{\sigma_0}  \dfrac{g_r g_\varphi}{\overline{\Sigma}^2}   \left( \dfrac{ \Sigma}{\sigma_0 \overline{\Sigma} } \right)^{s-1} \dif \Sigma.
\label{eq:algrav}
\end{equation}
Here, we made the important assumption that the $\Sigma$-PDF remains self-similar and that there is no physical scale at which the power-law breaks up. Indeed it is well known that self-gravity generates power-law density PDF which in the context of molecular clouds for instance tends to be proportional to $\rho^{-3/2}$ (see e.g. \cite{2011ApJ...727L..20K}, \cite{2018A&A...611A..88L}).
If we approximate $g_\varphi$ and $g_r$ as in Equations \eqref{eq:approx_gphi} and \eqref{eq:approx_gr}, Equation~\eqref{eq:algrav} becomes
\begin{align}
\nonumber
\alpha_{\mathrm{grav}} 
&= \dfrac{2}{3} \dfrac{ 2 \pi G h \varepsilon_r  \varepsilon_\varphi}{c_s^2 } \int _{ \sigma_0 \overline{\Sigma}} ^\infty  \sigma_0 P_0 \left( \dfrac{ \Sigma }{\sigma_0 \overline{\Sigma} } \right)^{s+1} \dif \Sigma \\
&=  - \dfrac{2}{3} \dfrac{ 2 \pi G \overline{\Sigma} h  \varepsilon_r  \varepsilon_\varphi }{ c_s^2 }    \dfrac{\sigma_0^2 P_0 }{s + 2}.
\label{eq:algrav2}
\end{align}
On the other hand, we know from a thermodynamical equilibrium  that the parameter $\alpha$, linked to the heating process, is directly related
to the coefficient $\beta$, which describes the cooling \citep{2001ApJ...553..174G}. In our case, Equation~\eqref{eq:alphabeta} writes
\begin{equation}
\label{eq:albeta}
\alpha = \dfrac{2}{5} \dfrac{1 }{\beta}.
\end{equation}
The parameter $\alpha$ appearing in Equation~\eqref{eq:albeta} is the sum of the gravitational contribution from Equation \eqref{eq:algrav2} and the contribution of the Reynolds stress (accounting for density and velocity fluctuations). We neglect the second for simplicity, which seems reasonable for such massive disc (see Figure~5 of \cite{2004MNRAS.351..630L}).  Therefore combining Equations \eqref{eq:algrav2} and \eqref{eq:albeta}, we get
\begin{equation}
s = - \dfrac{5}{3} \dfrac{2 \pi G \overline{\Sigma}   h \varepsilon_r  \varepsilon_\varphi}{c_s^2}\ \sigma_0^2  P_0\ \beta - 2.
\label{eq:algrav3}
\end{equation}
In our self-gravitating assumption, the scale height $h$ is written as
\begin{equation}
h = \dfrac{c_s^2}{\pi G \overline{\Sigma}}
\label{eq:h_sg}
\end{equation}
We thus retrieve the following linear relationship:
\begin{equation}
s = - \dfrac{10}{3} \varepsilon_r  \varepsilon_\varphi \sigma_0^2 P_0\ \beta - 2.
\label{eq:slin_an}
\end{equation}
We compare it to the empirical relation \eqref{eq:linship}. The intercept value of $-2$ is in good agreement with the value of $b_s~=~-1.6$ found in our simulations. Both the approximations used for the model and the statistical variations in the simulations do not allow for more precision. 
The value of the linear coefficient $a_s~=~0.2$ in our simulation and the fact that $\sigma_0^2P_0~\simeq~2$ suggest that $\varepsilon_r  \varepsilon_\varphi \simeq 0.025$. We would like to stress again that the assumption of only considering the contribution of the power-law zone of the PDF we made to write Equation~\eqref{eq:algrav} is crucial to end up with a linear relationship. 
This also true for our choice of $\sigma_0$ (see Figure~\ref{fig:sigma0_P0}), which suggests that this value is a threshold for the development of efficient gravitational heating. 

However, the assumption of a gas dominated radial gravity field (Equation~\eqref{eq:approx_gr}) is not as important. In the other extreme case where the radial component $g_r$ is dominated by the stellar contribution ($g_r \simeq GM_\star / r^2$), we get
\begin{equation}
s  = - \dfrac{5}{3} \dfrac{G M_\star h \varepsilon_\varphi} {c_s^2 r^2}\  \sigma_0 P_0\ \beta - 1.
\label{eq:algrav4}
\end{equation}
We can write the scale height as
\begin{equation}
h^2 = \dfrac{r^3 c_s^2}{G M_\star},
\end{equation}
and we finally get
\begin{equation}
s  = - \dfrac{5}{3} \dfrac{r}{h}\  \varepsilon_\varphi  \sigma_0 P_0\ \beta - 1.
\end{equation}
which is also a linear relationship given that $r / h$ is constant. 

\subsection{The Σ-PDF as a predictive tool}
\label{subsec:predictive}

Our characterization of  $\Sigma$-PDF of $\beta$-cooled self-gravitating discs may be used as a predictive tool to estimate the amount of fragmentation of a disc from the value of $\beta$.
We empirically found that the overdense region of the $\Sigma$-PDF can be described by a power-law (Equation~\eqref{eq:pwlw}).
We also found that the relation between the slope $s$ and $\beta$ is linear (Equation~\eqref{eq:linship}).
From these two equations, we can derive the following expression for the overdensity region of the $\Sigma$-PDF
\begin{equation}
    \mathcal{P}_{\beta}\left(\log(\sigma)\right) =  P_0 \left(\dfrac{\sigma}{\sigma_0}\right)^{a_s \beta + b_s}.
\end{equation}

The proportion of the disc where the fluctuation of column density overpass a given threshold $\sigma_{\mathrm{frag}}$  can be then estimated by the integral 
\begin{align}
\label{eq:intcrit}
P_{\mathrm{frag}}(\beta) &= \int^\infty_{\log\sigma_{\mathrm{frag}}} \mathcal{P}_\beta(\log\sigma) \dif\log\sigma \\
 &= \int^\infty_{\sigma_{\mathrm{frag}}} \mathcal{P}_\beta(\log\sigma)
 \dfrac{\dif\sigma}{\sigma} \\
 &= \dfrac{P_0}{\sigma_0}\int^\infty_{\sigma_{\mathrm{frag}}} \left(\dfrac{\sigma}{\sigma_0}\right)^{s - 1}
 \dif\sigma.
\end{align}
Since for $\beta > 0, s < 0$,
\begin{align}
\nonumber
P_{\mathrm{frag}}(\beta) 
&=  - \dfrac{P_0}{s} \left(\dfrac{\sigma_{\mathrm{frag}}}{\sigma_0}\right)^{s} \\
&=   - \dfrac{P_0}{a_s \beta + b_s} \left(\dfrac{\sigma_{\mathrm{frag}}}{\sigma_0}\right)^{a_s \beta + b_s}.
\label{eq:pfrag}
\end{align} 

\begin{figure*}
\centering
\includegraphics[width=\textwidth]{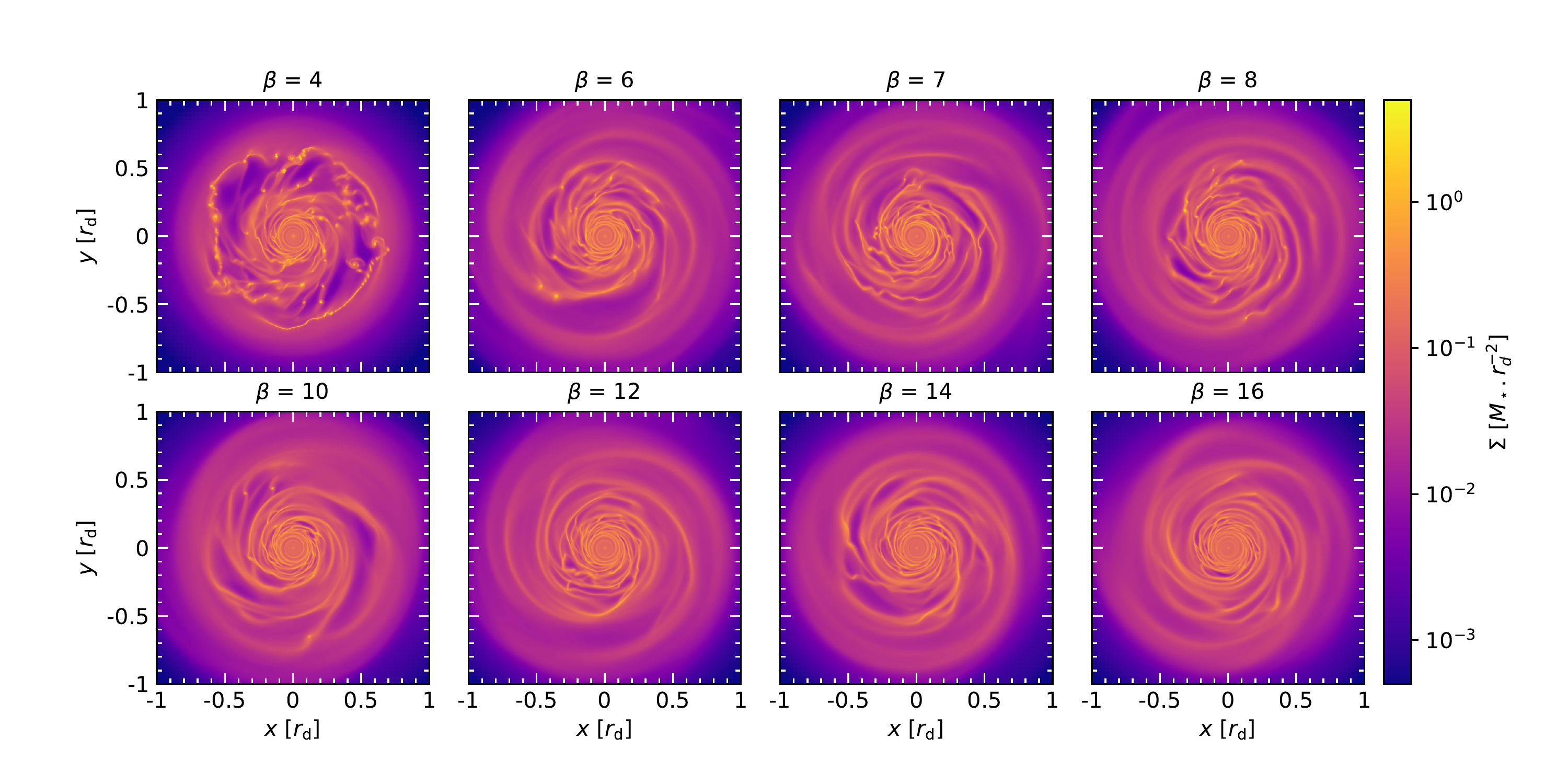} 

\caption{Column density maps for the group \textsc{JR13\_TIC}. Fragments form within the FSP.}\label{fig:coldens_jr13}

\end{figure*}

With Definition 1,  $\sigma_{\mathrm{frag}} = 30$. If the corresponding surface
$
S_{\mathrm{frag}} =  P_{\mathrm{frag}}(\beta) \pi r_d^2
$
is lower than the minimal resolved surface, no fragment can be seen, but they will appear when we increase the resolution. Note that Relation \eqref{eq:pfrag} only quantify the probability of forming dense clumps of gas but says nothing on their boundedness and time of survival. Another important parameter is also by how many cells the surface $S_{\mathrm{frag}}$ is resolved.
In Figure~\ref{fig:sfrag}, we compare the prediction made by this empirical model and the values we measured in the simulations. The results indicates that $S_{\mathrm{frag}}$ must be resolved by at least approximately 40 cells before we begin to observe fragments.

The important point here is that the characterization of the $\Sigma$-PDF gives a way to estimate the proportion of the surface of a disc that undergoes fragmentation. This enables to bypass the ill-determined $\beta_\mathrm{crit}$ as a criterion for fragmentation. 
The relationship between the index of the power-law $s$ with $\beta$ is well converged for $l_{\max} \geq 12$ as can be seen in Figure~\ref{fig:kappabeta}.

\section{First step of the gravitational cascade: the formation of a filamentary spiral pattern}
\label{sec:stepone}

\begin{figure}
\centering
\includegraphics[width=0.47 \textwidth]{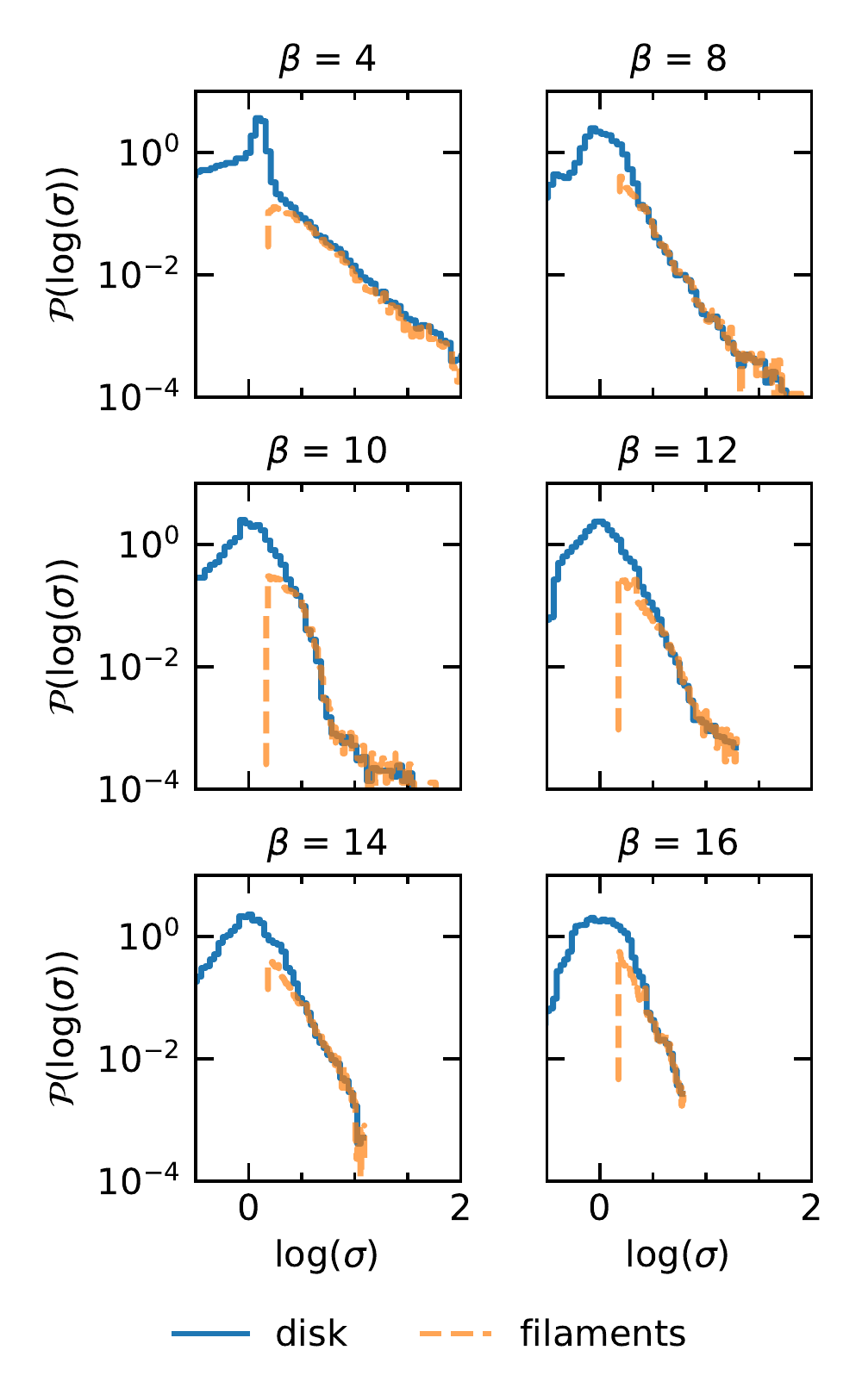} 

\caption{$\Sigma$-PDF at approximately 4.5 ORPs for the group \textsc{JR13\_TIC}. The blue line in the PDF for the disc while the orange dashed line is the PDF restricted to the extracted filaments (see the text). The two PDFs correspond in the overdense part.}\label{fig:pdf_fil}

\end{figure}

\begin{figure}
\centering
\includegraphics[width=0.45 \textwidth]{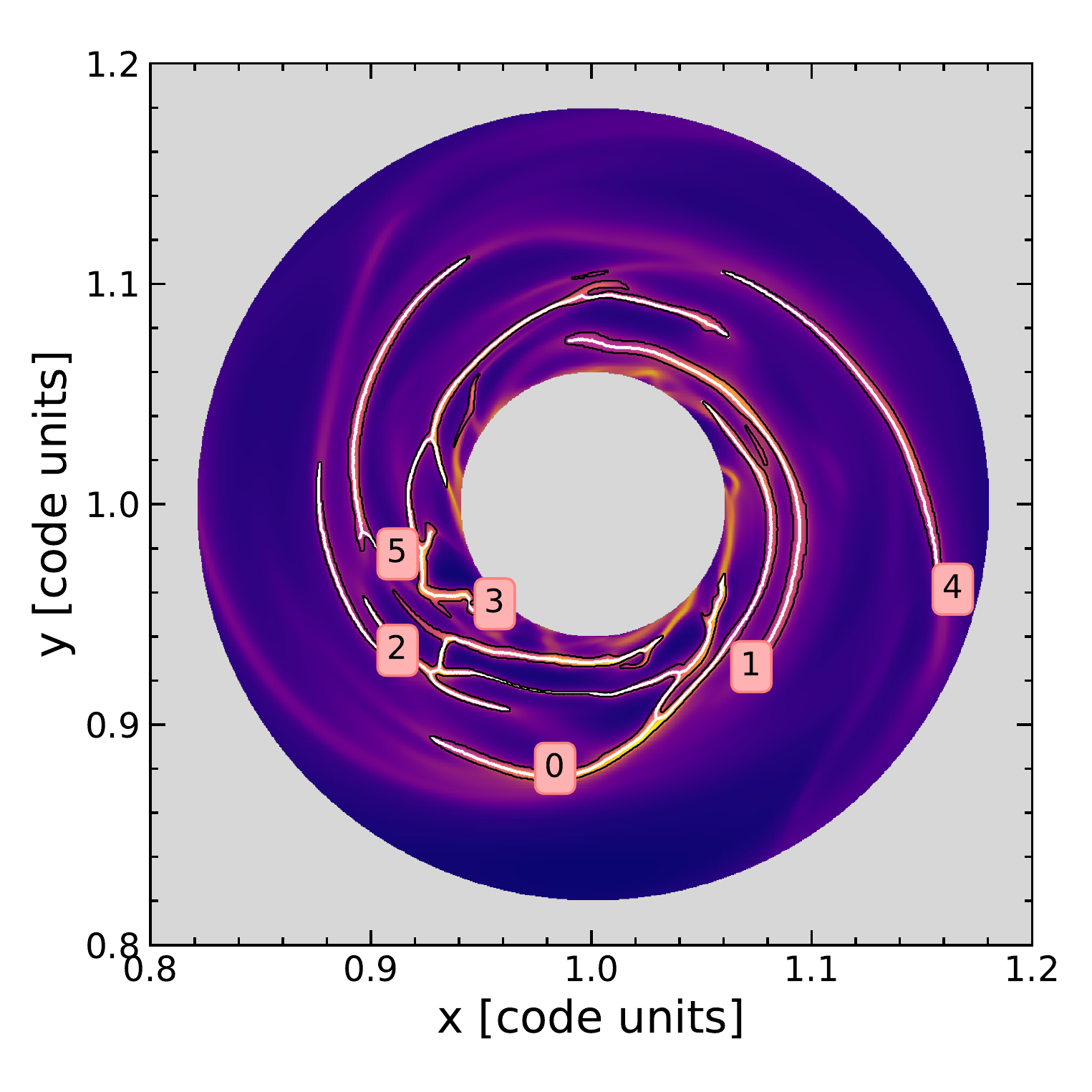} 

\caption{Example of filament extraction with \textsc{filfinder}, for the simulation $\beta = 6$ of the group \textsc{JR13\_TIC}. The brighter pixels correspond to the extracted filaments and the white line is the skeleton of individual filaments. The inner ($r < 0.04$) and outer ($r > 0.18$) regions of the disc are masked out.
Some filaments are not extracted by the algorithm because there are below the threshold.     }\label{fig:fil_mask}

\end{figure}

\begin{figure}
\centering
\begin{tikzpicture}[x=1.15pt,y=1.01pt,yscale=-1,xscale=1.]

\draw  [color={rgb, 255:red, 245; green, 166; blue, 35 }  ,draw opacity=1 ] (100.83,239.23) .. controls (100.83,234.95) and (104.9,231.48) .. (109.92,231.48) .. controls (114.93,231.48) and (119,234.95) .. (119,239.23) .. controls (119,243.51) and (114.93,246.99) .. (109.92,246.99) .. controls (104.9,246.99) and (100.83,243.51) .. (100.83,239.23) -- cycle ; \draw  [color={rgb, 255:red, 245; green, 166; blue, 35 }  ,draw opacity=1 ] (100.83,239.23) -- (119,239.23) ; \draw  [color={rgb, 255:red, 245; green, 166; blue, 35 }  ,draw opacity=1 ] (109.92,231.48) -- (109.92,246.99) ;
\draw  [color={rgb, 255:red, 0; green, 0; blue, 0 }  ,draw opacity=0.41 ] (46.71,174.09) .. controls (64.54,166.06) and (85.24,161.36) .. (107.38,161.16) .. controls (144.66,160.8) and (178.22,173.25) .. (200.96,193.17) -- (157.37,222.44) .. controls (144.35,215.32) and (127.01,211.06) .. (108.01,211.24) .. controls (97.38,211.34) and (87.29,212.82) .. (78.22,215.4) -- cycle ;
\draw [color={rgb, 255:red, 155; green, 155; blue, 155 }  ,draw opacity=1 ][fill={rgb, 255:red, 74; green, 74; blue, 74 }  ,fill opacity=1 ] [dash pattern={on 0.84pt off 2.51pt}]  (110,175.77) -- (110,240) ;
\draw  (60,120.06) -- (160.67,120.06)(70.44,30) -- (70.44,130) (153.67,115.06) -- (160.67,120.06) -- (153.67,125.06) (65.44,37) -- (70.44,30) -- (75.44,37)  ;
\draw  [fill={rgb, 255:red, 74; green, 144; blue, 226 }  ,fill opacity=1 ] (110.17,197.8) .. controls (109.16,197.8) and (108.33,197.1) .. (108.32,196.23) .. controls (108.31,195.36) and (109.12,194.65) .. (110.14,194.64) .. controls (111.15,194.63) and (111.98,195.33) .. (111.99,196.2) .. controls (112,197.07) and (111.19,197.79) .. (110.17,197.8) -- cycle ;
\draw  [fill={rgb, 255:red, 208; green, 2; blue, 27 }  ,fill opacity=1 ] (110.17,185.49) .. controls (109.16,185.49) and (108.33,184.79) .. (108.32,183.92) .. controls (108.31,183.05) and (109.12,182.34) .. (110.14,182.33) .. controls (111.15,182.32) and (111.98,183.02) .. (111.99,183.89) .. controls (112,184.76) and (111.19,185.48) .. (110.17,185.49) -- cycle ;
\draw   (83.11,167.18) -- (131.11,167.18) -- (131.11,208.45) -- (83.11,208.45) -- cycle ;
\draw    (110,167.18) -- (110,152) ;
\draw [shift={(110,150)}, rotate = 450] [color={rgb, 255:red, 0; green, 0; blue, 0 }  ][line width=0.75]    (10.93,-3.29) .. controls (6.95,-1.4) and (3.31,-0.3) .. (0,0) .. controls (3.31,0.3) and (6.95,1.4) .. (10.93,3.29)   ;
\draw [color={rgb, 255:red, 155; green, 155; blue, 155 }  ,draw opacity=1 ][fill={rgb, 255:red, 74; green, 74; blue, 74 }  ,fill opacity=1 ] [dash pattern={on 0.84pt off 2.51pt}]  (110,30) -- (110,138) ;
\draw [shift={(110,140)}, rotate = 270] [color={rgb, 255:red, 155; green, 155; blue, 155 }  ,draw opacity=1 ][line width=0.75]    (10.93,-4.9) .. controls (6.95,-2.3) and (3.31,-0.67) .. (0,0) .. controls (3.31,0.67) and (6.95,2.3) .. (10.93,4.9)   ;
\draw  [fill={rgb, 255:red, 74; green, 144; blue, 226 }  ,fill opacity=1 ] (110.77,101.65) .. controls (109.76,101.66) and (108.93,100.84) .. (108.92,99.83) .. controls (108.91,98.81) and (109.72,97.98) .. (110.74,97.97) .. controls (111.75,97.96) and (112.58,98.78) .. (112.59,99.79) .. controls (112.6,100.81) and (111.79,101.64) .. (110.77,101.65) -- cycle ;
\draw  [fill={rgb, 255:red, 208; green, 2; blue, 27 }  ,fill opacity=1 ] (110.77,62.32) .. controls (109.76,62.33) and (108.93,61.51) .. (108.92,60.49) .. controls (108.91,59.48) and (109.72,58.65) .. (110.74,58.64) .. controls (111.75,58.63) and (112.58,59.44) .. (112.59,60.46) .. controls (112.6,61.47) and (111.79,62.31) .. (110.77,62.32) -- cycle ;
\draw [color={rgb, 255:red, 208; green, 2; blue, 27 }  ,draw opacity=0.49 ]   (110,61) -- (110,79) ;
\draw [shift={(110,81)}, rotate = 270] [color={rgb, 255:red, 208; green, 2; blue, 27 }  ,draw opacity=0.49 ][line width=0.75]    (10.93,-3.29) .. controls (6.95,-1.4) and (3.31,-0.3) .. (0,0) .. controls (3.31,0.3) and (6.95,1.4) .. (10.93,3.29)   ;
\draw [color={rgb, 255:red, 126; green, 211; blue, 33 }  ,draw opacity=0.79 ]   (110,59) -- (110,41) ;
\draw [shift={(110,39)}, rotate = 450] [color={rgb, 255:red, 126; green, 211; blue, 33 }  ,draw opacity=0.79 ][line width=0.75]    (10.93,-3.29) .. controls (6.95,-1.4) and (3.31,-0.3) .. (0,0) .. controls (3.31,0.3) and (6.95,1.4) .. (10.93,3.29)   ;
\draw [color={rgb, 255:red, 189; green, 16; blue, 224 }  ,draw opacity=0.75 ] [dash pattern={on 4.5pt off 4.5pt}]  (109.89,58.28) -- (109.99,22) ;
\draw [shift={(110,20)}, rotate = 450.17] [color={rgb, 255:red, 189; green, 16; blue, 224 }  ,draw opacity=0.75 ][line width=0.75]    (10.93,-3.29) .. controls (6.95,-1.4) and (3.31,-0.3) .. (0,0) .. controls (3.31,0.3) and (6.95,1.4) .. (10.93,3.29)   ;
\draw   (50,10) -- (180,10) -- (180,150) -- (50,150) -- cycle ;

\draw (121,225) node [anchor=north west][inner sep=0.75pt]  [font=\normalsize,color={rgb, 255:red, 74; green, 74; blue, 74 }  ,opacity=1 ] [align=left] {Central object};
\draw (140,185) node [anchor=north west][inner sep=0.75pt]  [font=\normalsize,color={rgb, 255:red, 74; green, 74; blue, 74 }  ,opacity=1 ] [align=left] {Filament};
\draw (92.9,193.5) node [anchor=north west][inner sep=0.75pt]  [font=\normalsize,color={rgb, 255:red, 74; green, 144; blue, 226 }  ,opacity=1 ]  {$r_{0}$};
\draw (113.33,186.82) node [anchor=north west][inner sep=0.75pt]  [font=\normalsize,color={rgb, 255:red, 74; green, 144; blue, 226 }  ,opacity=1 ]  {$C$};
\draw (92.47,177.19) node [anchor=north west][inner sep=0.75pt]  [font=\normalsize,color={rgb, 255:red, 208; green, 2; blue, 27 }  ,opacity=1 ]  {$r_{F}$};
\draw (114.4,173.22) node [anchor=north west][inner sep=0.75pt]  [font=\normalsize,color={rgb, 255:red, 208; green, 2; blue, 27 }  ,opacity=1 ]  {$F$};
\draw (77,22.4) node [anchor=north west][inner sep=0.75pt]  [font=\normalsize]  {$y$};
\draw (161,102.4) node [anchor=north west][inner sep=0.75pt]  [font=\normalsize]  {$x$};
\draw (93.5,97.73) node [anchor=north west][inner sep=0.75pt]  [font=\normalsize,color={rgb, 255:red, 74; green, 144; blue, 226 }  ,opacity=1 ]  {$r_{0}$};
\draw (113.93,89.79) node [anchor=north west][inner sep=0.75pt]  [font=\normalsize,color={rgb, 255:red, 74; green, 144; blue, 226 }  ,opacity=1 ]  {$C$};
\draw (93.07,53.74) node [anchor=north west][inner sep=0.75pt]  [font=\normalsize,color={rgb, 255:red, 208; green, 2; blue, 27 }  ,opacity=1 ]  {$r_{F}$};
\draw (115,48.96) node [anchor=north west][inner sep=0.75pt]  [font=\normalsize,color={rgb, 255:red, 208; green, 2; blue, 27 }  ,opacity=1 ]  {$F$};
\draw (112,73.4) node [anchor=north west][inner sep=0.75pt]  [font=\scriptsize,color={rgb, 255:red, 208; green, 2; blue, 27 }  ,opacity=1 ]  {$g_{\mathrm{fil}}$};
\draw (121,131) node [anchor=north west][inner sep=0.75pt]  [font=\footnotesize,color={rgb, 255:red, 74; green, 74; blue, 74 }  ,opacity=1 ] [align=left] {towards the \\central object};
\draw (110.33,14.07) node [anchor=north west][inner sep=0.75pt]  [font=\scriptsize,color={rgb, 255:red, 189; green, 16; blue, 224 }  ,opacity=1 ]  {$R_{\mathrm{fil}}$};
\draw (88,36.4) node [anchor=north west][inner sep=0.75pt]  [font=\scriptsize,color={rgb, 255:red, 126; green, 160; blue, 33 }  ,opacity=1 ]  {$f_{p,\mathrm{fil}}$};
\draw    (129,10) -- (180,10) -- (180,25) -- (129,25) -- cycle  ;
\draw (132,14) node [anchor=north west][inner sep=0.75pt]  [font=\footnotesize] [align=left] {shearing box};

\end{tikzpicture}

\caption{Schematic view of a filament.}\label{fig:schema_fil}

\end{figure}
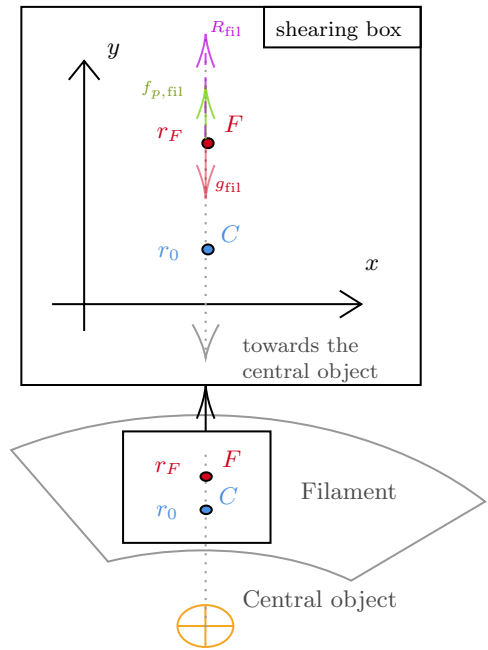

\begin{figure*}
\centering
\includegraphics[width=  \textwidth]{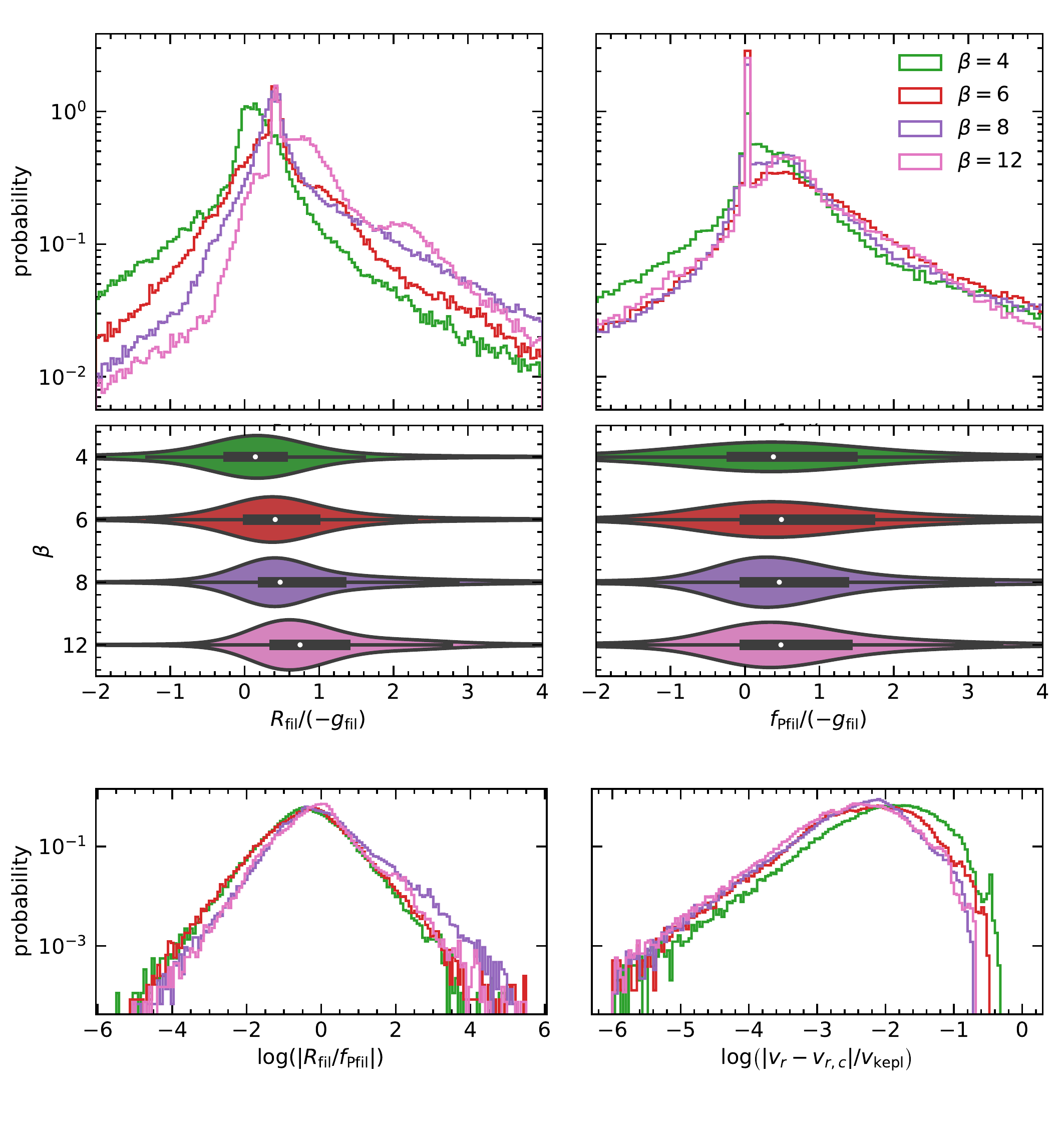} 

\caption{Ratio of forces within filaments. The notations are explained in section \ref{subsubsec:equilibrium} and in Table~\ref{tbl:notation}. All quantities are computed from filaments extracted from a mid-plane slice of the simulation of the group \textsc{JR13} at the same time as in  Figure~\ref{fig:coldens_jr13}. \textbf{Top left}: PDF of the ratio of the centrifugal force over gravity. The distribution shifts to right towards 1 when $\beta$ increases, meaning that rotational tends to match gravity.
The violin plot beneath displays a kernel density estimate of the distribution for each $\beta$. The median is pictured with a white circle and half of the distribution is 
within the black box.
\textbf{Top right}: PDF of the ratio of pressure force over gravity. The pressure force plays an important role in the stability of the filaments but $\beta$ has only a small influence on it.
\textbf{Bottom left}: Ratio of centrifugal force over pressure force. The distribution shifts to the right as $\beta$ increase, another hint that the importance of rotational support increases as the cooling becomes less efficient. \textbf{Bottom right}: PDF of the normalized difference between the radial velocity within the filaments and the radial velocity at the centre of the filaments. Almost everywhere within the filament the radial velocity is negligible with respect to the Keplerian speed (the filament are close to be at equilibrium radially). 
The probability of a radial collapse (given by the right tail of the distribution) is higher for low value of $\beta$. }
\label{fig:ratio_pdf}

\end{figure*}

\begin{figure}
\centering

\includegraphics[width= 0.48 \textwidth]{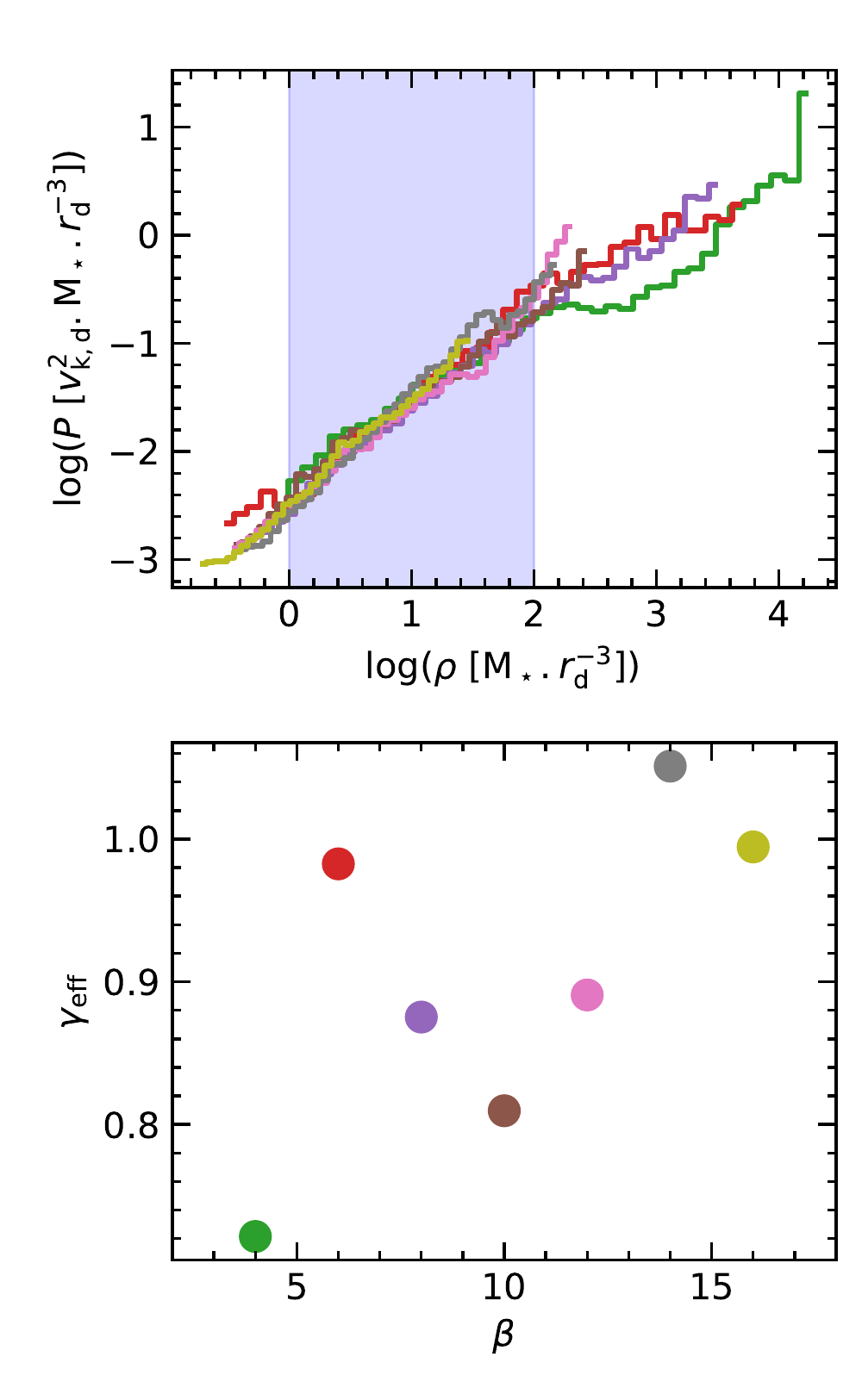} 

\caption{Pressure-density relation in the filaments. The top panel shows the averaged value of the logarithm of the pressure in logarithmic bins of density. The slope of this curve (bottom panel) is the adiabatic effective index $\gamma_{\text{eff}}$ within the filament. It was computed by fitting the above curves between 0 and 2 (blue shaded region) to exclude fragments. It is slightly lower than 1 (around 0.8), meaning that the filaments are close to isothermality. The data are computed from filaments extracted from the mid-plane slice of simulation of the group \textsc{JR13} (same times as in Figure~\ref{fig:coldens_jr13}).}
\label{fig:P}

\end{figure}

\begin{figure*}
\centering
\includegraphics[width= 0.94\textwidth]{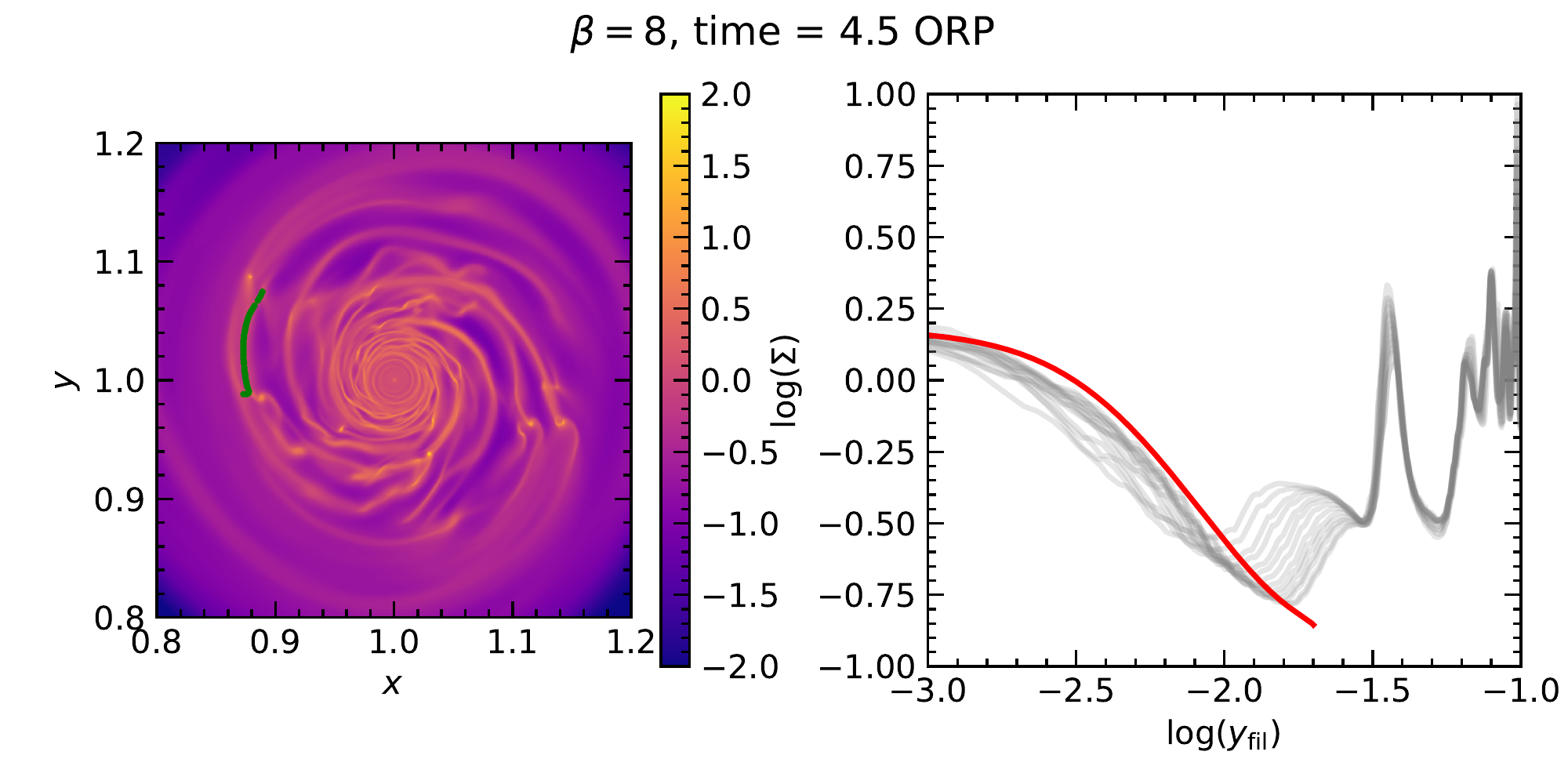} 

\caption{The grey lines on the right-hand panel are profiles of column density from the green filament section represented on the left-hand panel. The filament is extracted from the simulation $\beta = 8$ of the group \textsc{JR13} at 4.5 ORPs. The red line in the right-hand panel is given by the analytical model of the section \ref{subsec:analytical} with $K = 1.38$, the other parameters $\Sigma_0$, $\Omega_0$ and $c_s$ being taken from the simulation. Distances and column densities are in code units.}
\label{fig:fil_prof}

\end{figure*}

From Figures \ref{fig:coldens} and \ref{fig:coldens_jr13}, we can see that for all the values of $\beta$ we considered, the disc develops a Filamentary Spiral Pattern (FSP).
The dense gas is mainly found in the filaments, so we expect that the shape of the overdense region of the $\Sigma$-PDF results from their $\Sigma$-PDF (see Figure~\ref{fig:pdf_fil}). 
Furthermore, when the disc fragments, fragments appear within the FSP. 
From the study of these filaments, we devised a two-step scenario for the condensation of gas within the disc. 
The first step, presented below, is the formation of almost radially stable filaments. The second one, presented in the section \ref{sec:steptwo}, is the collapse of the gas alongside these filaments.

\subsection{The properties of the filaments in our simulation}
\label{subsec:fil_prop}

\subsubsection{Extraction}
\label{subsubsec:extraction}
We use the package \textsc{filfinder} \citep{2015MNRAS.452.3435K} to extract filaments from a map of column density (Figure~\ref{fig:fil_mask}). 
This package detects filaments with an adaptive threshold from a flattened image. 
This allows to address the structural dynamic range of column density of the disc and the high overdensities due to fragments.
Figure~\ref{fig:pdf_fil} shows that the \mbox{$\Sigma$-PDF} restricted to the extracted FSP reproduce the shape of the overdense region of the $\Sigma$-PDF of the whole disc. It also reproduces the slope and its variation with $\beta$.
This good agreement together with the probability analysis (Figure~\ref{fig:sfrag}) and the visual impression that fragments form inside the FSP (Figures \ref{fig:coldens} and \ref{fig:coldens_jr13}) stresses the need to understand what drives their evolution.

\subsubsection{Equilibrium in the filaments}
\label{subsubsec:equilibrium}

To better understand the shape of the filaments we analyse their support against self-gravity. The radial projection of Euler's equation in cylindrical coordinates for a given point in the disc writes
\begin{equation}
\label{eq:euler}
-\dfrac{GM_\star}{r^2} + g_{\mathrm{gas}} +
 \dfrac{v_\varphi^2}{r} -
 \dfrac{\partial_r P}{\rho} = \partial_t v_r + v_r\partial_r v_r,
\end{equation}
where $r$ and $\varphi$ are respectively the radial and azimuthal coordinates, $P$ is the pressure and  $g_{\mathrm{gas}}$ the gravitational field due to the gas.

Let's consider a filament like in the cartoon of Figure~\ref{fig:schema_fil}. We assume that the filament is locally perpendicular to the radial direction.
We consider a given point $F$ within a filament at a distance $r_\mathrm{F}$ from the star and the corresponding point $C$ at the centre of the filament in the same radial direction with a radius $r_0$.
By subtracting Equation~\eqref{eq:euler} evaluated at $F$ and at $C$ we get
\begin{multline}
\label{eq:euler_subs}
g_{\mathrm{gas, F}} - g_{\mathrm{gas, C}} 
+ \dfrac{\partial_r P_F}{\rho_F} - \dfrac{\partial_r P_C}{\rho_C} 
+  \dfrac{v_{\varphi\mathrm{, F}}^2}{r_F} -  \dfrac{v_{\varphi\mathrm{, C}}^2}{r_C}
-  \dfrac{v_\mathrm{k, F}^2}{r_F} +  \dfrac{v_\mathrm{k, C}^2}{r_C} \\
= \partial_t v_{r,\mathrm{F}} + v_{r,\mathrm{F}}\partial_r v_{r,\mathrm{F}} - \left( \partial_t v_{r,\mathrm{C}} + v_{r,\mathrm{C}}\partial_r v_{r,\mathrm{C}} \right),
\end{multline}
where $v_\mathrm{k} = -\sqrt{\frac{GM_\star}{r}}$ is the Keplerian speed.

We assume that the filament is at equilibrium.
This assumption is sustained by the fact that once formed, filaments last for several orbital periods and seems neither to expand nor to retract.
With this assumption, the right-hand term $Dv_{r, \mathrm{fil}} = \partial_t v_{r,\mathrm{F}} + v_{r,\mathrm{F}}\partial_r v_{r,\mathrm{F}} - \left( \partial_t v_{r,\mathrm{C}} + v_{r,\mathrm{C}}\partial_r v_{r,\mathrm{C}} \right)$ is nul. The validity of this assumption is supported by the bottom right-hand panel of Figure~\ref{fig:ratio_pdf}.
We define:
\begin{equation}
    g_{\mathrm{fil}} = g_{\mathrm{gas, F}} - g_{\mathrm{gas, C}},
\end{equation}
the self-gravity of the filament,
\begin{equation}
   R_{\mathrm{fil}} =  \left( \dfrac{v_{\varphi\mathrm{, F}}^2}{r_F} - \dfrac{v_\mathrm{k, F}^2}{r_F}  \right)
- \left(  \dfrac{v_{\varphi\mathrm{, C}}^2}{r_C} -  \dfrac{v_\mathrm{k, C}^2}{r_C} \right),
\end{equation}
the support/collapse term due to differential rotation, and
\begin{equation}
    f_{\mathrm{Pfil}} = \dfrac{\partial_r P_F}{\rho_F} - \dfrac{\partial_r P_C}{\rho_C},
\end{equation}
the thermal support. Assuming that the geometrical centre of the filament coincides with the density and pressure maximum of the filament, the second term  ${\partial_r P_C}/{\rho_C}$ is equal to zero. In reality the filaments can be slightly asymmetrical and a little correction would be needed, but is neglected in this study. The equilibrium in the filaments is written as
\begin{equation}
    \label{eq:fil_equ}
    g_{\mathrm{fil}} +  f_{\mathrm{Pfil}} + R_{\mathrm{fil}}  = 0.
\end{equation}

We first investigate whether a stronger $\beta$ (that is a less efficient cooling) results in a higher effective polytropic index $\gamma_{\mathrm{eff}}$ such as $P = \rho^{\gamma_{\mathrm{eff}}}$. 
This may result in a stronger thermal support in the high $\beta$ case and explain a steepest $\Sigma$-PDF. Figure~\ref{fig:P} shows that there is no evidence of such a relationship between $\gamma_{\mathrm{eff}}$ in the filaments and $\beta$.

 Secondly, we investigate the various support, thermal and rotational within the filaments. Figure~\ref{fig:ratio_pdf} shows 
 PDFs of the ratio of $R_{\mathrm{fil}}$, $g_{\mathrm{fil}}$ and $ f_{\mathrm{Pfil}}$ for several values of $\beta$.  The top right-hand and bottom left-hand panels show that the thermal support within the filament should be taken into account but does not depends so much on $\beta$. 
In simulations with higher $\beta$, the support provided by the differential rotation is stronger. For instance, on Figure~\ref{fig:ratio_pdf} top left-hand panel, in more of 50 \% of the surface of filaments $R_{\mathrm{fil}} / \vert g_{\mathrm{fil}} \vert$ is comprised between $0$ and $1$ for $\beta = 4$, and between $0.5$ and
$1.5$ for $\beta = 12$ as the distribution shifts to the right as $\beta$ increases.
This can be explained by the fact that simulation with a lower $\beta$ (and thus more efficient cooling) are subject to gravitational instabilities that generates turbulent motions within the disc \citep{2001ApJ...553..174G}. This turbulent motions dissipate angular momentum and thus reduce the rotational support. 
The role of the drop of rotational support due to viscous motions was previously stressed by \cite{2016ApJ...824...91L}. 

Our statistical measurements demonstrate the importance of the differential rotation in the processes of shaping the filaments and their PDF and thus in the fragmentation process. It would be interesting if we could use these finding to build a model explaining the actual shape of the filament as in the example of the Figure~\ref{fig:fil_prof}. This is the object of the next section.

\subsection{Analytical model for the shape of the filaments}
\label{subsec:analytical}

To assess the results obtained in the numerical simulations (section \ref{subsec:fil_prop}), we seek for an analytical 
model to describe the FSP that forms in the disc. This self-gravitating structures is supported by thermal pressure and rotation. 

\subsubsection{Analytical framework and assumptions}
\label{subsubsec:framework}
For the purpose of building a analytical model for the filaments in the disc, we make several simplifying assumptions:
\begin{enumerate}
\item we make the shearing box approximation,
\item we consider that the filaments are at mechanical equilibrium,
\item we assume that the gas is locally isothermal,
\item we assume that the filaments are thin.

\end{enumerate}
As we shall see, some of these assumptions can be disputed, even in the simplified view of the simulations discussed above. 
However, they provide a simple analytical framework that permits to illustrate how the interplay between gravity, rotation, pressure and energy dissipation shapes the filaments and the $\Sigma$-PDF.

\paragraph*{Shearing box approximation and mechanical equilibrium:}

As in section \ref{subsec:fil_prop}, we focus on the equilibrium of filaments. Like previous studies \citep{2001ApJ...553..174G, 2012MNRAS.421.3286P, 2017A&A...606A..70K} we use a local model (or shearing box). \cite{2004MNRAS.351..630L, 2005MNRAS.358.1489L} investigated the validity of such a model for self-gravitating discs and found no evidence of global wave energy transport, allowing for a local treatment of the energy dissipation.
We consider a small region within a filament at a radius $r_0$ comoving with the disc at the angular speed $\Omega_0$ (see Figure~\ref{fig:schema_fil}). We introduce the local coordinates $x = r -  r_0$ and $y = r_0 \left( \varphi - \Omega t \right)$.  We assume that the disc is Keplerian and that the filament is at mechanical equilibrium, that is  $D\bm{v}/Dt = 0$ and $v_y = 0$. We also assume that the filament is azimuthally uniform, that is $\partial_x = 0$. 
The equations of motion are expanded to the first order in $|y| /  r_0$:

\begin{equation}
    \label{eq:euler_shbox}
     g_\mathrm{fil} \bm{e_y} + 3 \Omega_0^2 y \bm{e_y} - \Omega_0^2 z \bm{e_z}  - 2  \Omega_0 \bm{e_z} \times \bm{v}  - \dfrac{1}{\rho}\nabla P = 0.
\end{equation}

\paragraph*{Isothermal behaviour:}

In our simulations, the adiabatic index of the gas is $\gamma~=~5/3$. However, the effective adiabatic index in the filaments is much lower with $\gamma_{\mathrm{eff}}~\sim~0.85$ (Figure~\ref{fig:P}). Filaments are not so far from being isothermal and assuming them to be so greatly simplifies the equations. In our model, 
\begin{equation}
    \label{eq:cs}
    P = c_s^2 \rho
\end{equation}
and thus
\begin{equation}
    \nabla P \cdot \bm{e_y} = c_s^2 \partial_y \rho
\end{equation}
where the sound speed $c_s$ is constant.

\paragraph*{Thin disc approximatiom:}

We assume that the disc is thin, that is $z \ll r$ and thus $\Omega_0^2 z \ll 1$ as well. 
For isothermal self-gravitating disc at vertical equilibrium, the scale height writes \citep{1965MNRAS.130..125G, 2016ARA&A..54..271K}:
\begin{equation}
\label{eq:h}
 h = \dfrac{ c_s}{\sqrt{2 \pi G \rho_c} },
\end{equation}
where $\rho_c$ is the density in the mid-plane. 
The column density can be written
\begin{equation}
\label{eq:coldens_thin}
\Sigma = 2 h \rho_c.
\end{equation}
By multiplying Equation~\eqref{eq:euler_shbox} by $2h$ and projecting in the $y$-direction we obtain
\begin{equation}
\label{eq:euler_int}
\Sigma \left( g_\mathrm{fil} + 3 \Omega_0 ^2 y -2 \Omega_0 v_x \right) - 2 c_s^2 \partial _y \Sigma = 0.
\end{equation}

\subsubsection{Transport of angular momentum}

An essential aspect in the problem under investigation is the transport of angular 
momentum through the $\alpha$-viscosity. The transport equation in the shearing
box approximation is
\begin{equation}
\label{eq:diff_mom}
\partial _t \Sigma v_x + \partial_y(\Sigma v_x v_y) = \partial_y (\nu \Sigma \partial_y v_x ),
\end{equation}
where $\nu = \alpha c_s h$ is the effective viscosity. Since we seek for stationary solutions, 
we simply require that $\nu \Sigma \partial_y v_x$ is equal to a constant. 
Combining Equations~(\ref{eq:h}) and~(\ref{eq:coldens_thin}), we see that $\Sigma \propto \sqrt{\rho_c}$;
therefore, $\nu \Sigma$ is a constant as long as the gas remains isothermal. We thus get 
\begin{equation}
\label{eq:diff_mom_eq}
v_x = K \Omega_0 y, 
\end{equation}
where $K$ is a dimensionless number. 
Combining this relation with Equation~\eqref{eq:euler_int}, we arrive to
\begin{equation}
\label{eq:euler_trans}
\Sigma \left( g_\mathrm{fil} + (3 -2 K) \Omega_0 ^2 y \right)  = 2 c_s^2 \partial _y \Sigma.
\end{equation}
This equation describes a filament that is at perfect mechanical equilibrium.

\subsubsection{Gravitational potential}

To get $ g_\mathrm{fil}$, we have to perform an integration through the filament. 
We assume that the filament is symmetrical with respect to the $x$-axis and 
goes from $y=-\Lambda$ to $y=\Lambda$. As in the thin disc geometry, the plane
$z=0$ is singular, we calculate the gravitational field at $z=\varepsilon$.
From the direct integration over $x$ and $y$ we get
\begin{align}
\label{eq:gfil}
 g_\mathrm{fil} (0,y,\varepsilon) &= G \int _{-\infty} ^\infty \int_{-\Lambda} ^\Lambda  \dfrac{\Sigma(x,y^\prime)(y - y^\prime)}{(x^2+{(y - y^\prime)}^2+\varepsilon^2)^{3/2}} \dif x \dif y^\prime \nonumber \\
&=
G I  \int_{-\Lambda} ^\Lambda  \dfrac{\Sigma(0,y^\prime) (y - y^\prime)}{{{(y - y^\prime)}^2+\varepsilon^2}} \dif y^\prime , 
\end{align}
where we took into account that $\Sigma$ is invariant along the $x$-axis and 
\begin{equation}
I = \int _{-\infty} ^\infty  \dfrac{\dif x^\prime}{(1 + {x^\prime}^2)^{3/2} } = 2.
\end{equation}

\subsubsection{Solution of the equation of motion}

For the purpose of solving the numerical 
problem we rewrite $ g_\mathrm{fil}$ as 
\begin{equation}
\label{eq:gfil_rw}
 g_\mathrm{fil} (0,y,\varepsilon) = 2G \int_0 ^\Lambda  \Sigma(0,y^\prime) D(y, y^\prime, \varepsilon) \dif y^\prime,
\end{equation}
where 
\begin{equation}
D(y, y^\prime, \varepsilon) = \dfrac{{y^\prime - y}}{{{(y - y^\prime)}^2+\varepsilon^2}} - 
\dfrac{y^\prime+y}{{(y^\prime+y)^2+\varepsilon^2}}.
\end{equation}
With this last expression, we see that Equation~\eqref{eq:euler_trans} is an integro-differential equation 
of the first order. It is determined by its boundary conditions.
We assume that $\partial_y\Sigma(0)~=~0$ and that $\Sigma(0)=\Sigma_0$ is given.
We start by normalizing Equation~\eqref{eq:euler_trans}. We write 
\begin{alignat}{5}
\label{eq:gfil_norm}
y &= y_\textsc{k}\widetilde{y}\;  &&{\rm with }&\; y_\textsc{k} &= \dfrac{\sqrt{2} c_s}{\sqrt{3 - 2 K} \Omega_0},  \\
\Sigma &= \Sigma_\textsc{k} \widetilde{\Sigma}\; &&{\rm with }&\;  \Sigma_\textsc{k} &= \dfrac{c_s^2}{ y_\textsc{k} G }, 
\end{alignat}
where it is assumed that $3 -2 K >0$ (i.e., centrifugal support exerts a support).
Equation~\eqref{eq:euler_trans} becomes
\begin{equation}
\label{eq:euler_norm}
 \dfrac{\partial_{\widetilde{y}}\widetilde{\Sigma}}{ \widetilde{\Sigma} } = \int_0 ^ {\widetilde{\Lambda}}  \widetilde{ \Sigma} (0,y^\prime) D(y^\prime, \widetilde{y}, \varepsilon) \dif y^\prime + \widetilde{y}.
\end{equation}

To generate a sequence of equilibrium we can simply increase the value of $\widetilde{\Sigma} (0)$ 
while imposing that the mass $\int _0 ^{\widetilde{\Lambda}} \widetilde{\Sigma}(y')  \dif y'$ remains invariant. 
In principle, this may have imposed to search for the corresponding 
$\widetilde{\Lambda} = y_\textsc{k} \Lambda$  for each value of $\widetilde{\Sigma} (0)$. In practice, it turns out that for $\widetilde{\Lambda} \simeq 1.5$ all solutions have approximately the same mass. 
To integrate Equation~\eqref{eq:euler_norm}, we use the python package \textsc{IDEsolver} \citep{Karpel2018}. 

\begin{figure}
\centering
\includegraphics[width=0.5\textwidth]{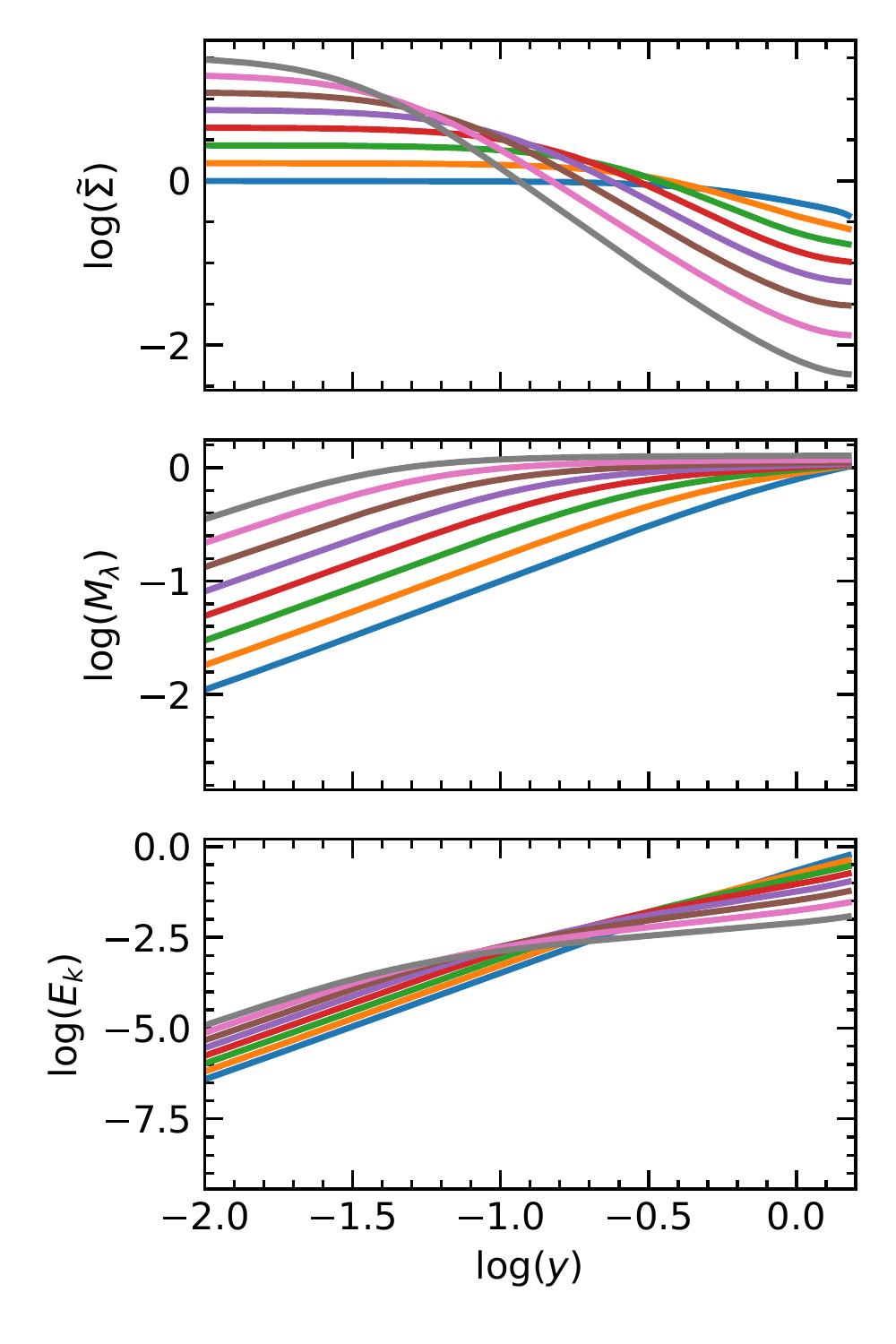}
\caption{Sequence of equilibrium of a rotating and self-gravitating filament. Each color corresponds to a central column density $\widetilde{\Sigma}(0)$ (see top panel).
Top panel portrays the column density, middle one the cumulative lineic mass perpendicularly to the filament and 
bottom one the rotation or kinetic energy per units of length.
 }
\label{fig:fil_eq}
\end{figure}

The results are displayed in Figure~\ref{fig:fil_eq}. 
The column density profiles (top panel) show that as the gas contracts along the $y$-axis
the filament becomes progressively more peaked. Middle panel reveals that indeed at $y=1.5$, 
all filament masses are almost identical while the kinetic energy decreases as it should along the sequence of equilibrium triggered by the viscous transport of angular momentum. 

Figure~\ref{fig:fil_prof} portrays a comparison between a solution of the analytical model and a series of filament profiles extracted from the simulation. 
The good agreement that can be seen confirms the proposed picture of a quasi-equilibrium contraction in the radial direction driven by the viscous transport of angular momentum. 
This is the first condensation step underwent by self-gravitating discs subject to $\beta$ cooling, leading to the formation of gas filaments.

\section{Second step of the gravitational cascade: collapse of the gas within filaments}
\label{sec:steptwo}

\begin{figure*}
\centering
\includegraphics[width= 0.9 \textwidth]{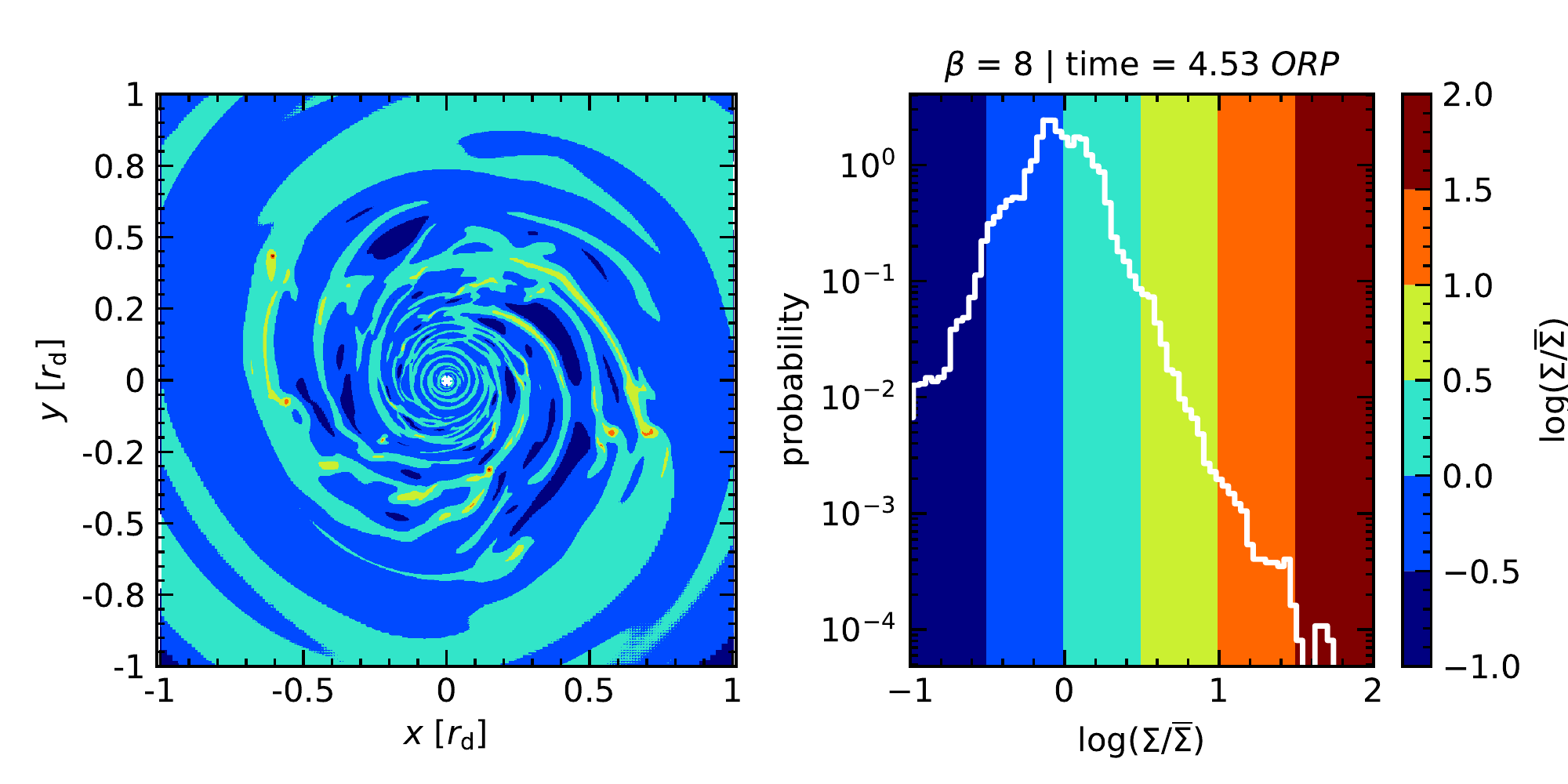} 

\caption{Map of fluctuations of column density $\sigma = \Sigma / \overline{\Sigma}$ (left) in logarithmic bins represented on the $\Sigma$-PDF (right) for $\beta = 8$. Dense parts of the PDF ($\log(\sigma) > 0.5$) are only present in some restricted regions of the filaments. }\label{fig:segmentation}

\end{figure*}

\begin{figure*}
\centering
\includegraphics[width= 0.95 \textwidth]{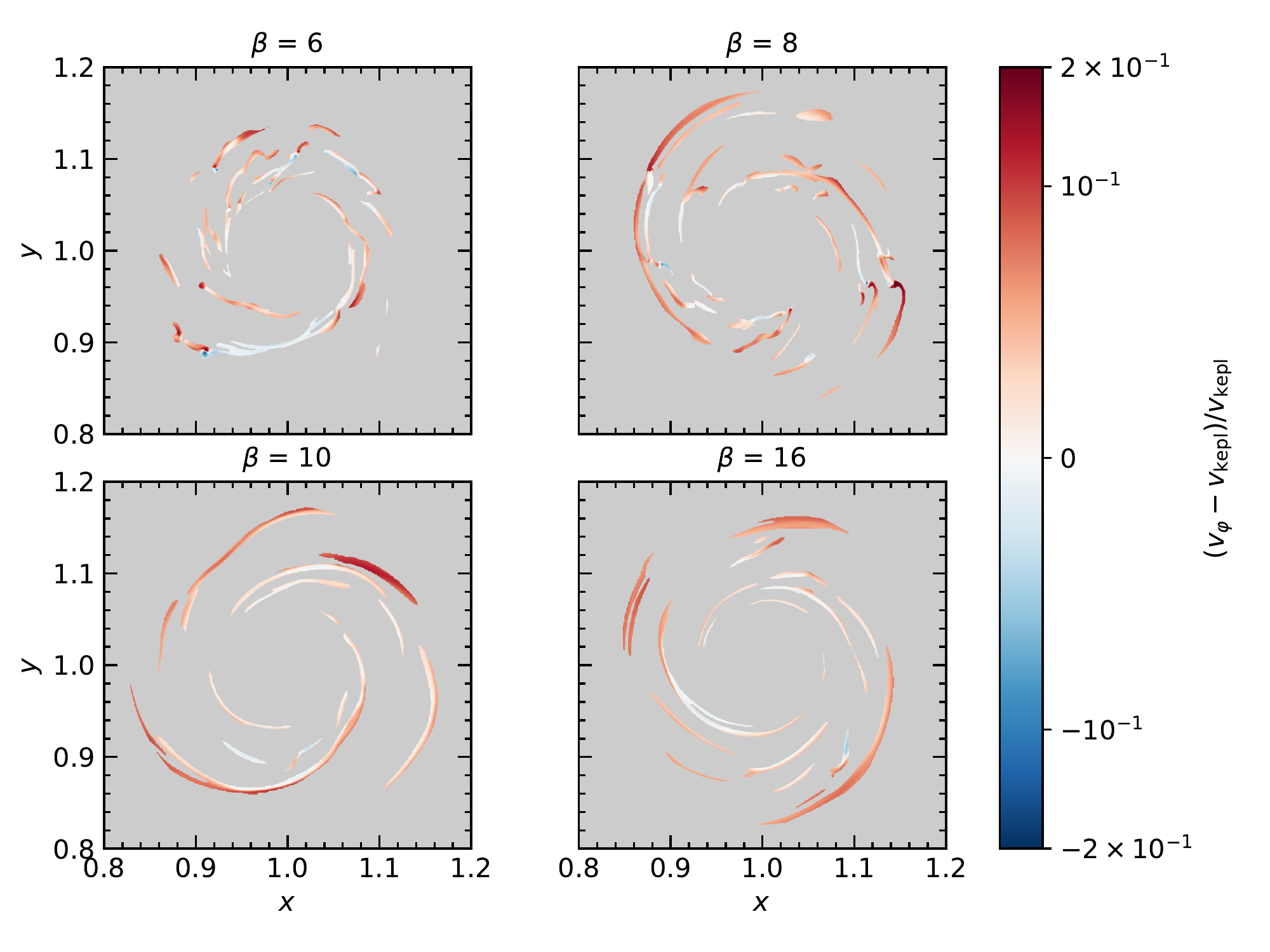} 

\caption{Azimuthal velocity $v_\varphi$ in the FSP relative to the 
Keplerian velocity $v_\mathrm{kepl}$ on mid-plane slice taken at about 4.5 ORPs.
Blue parts are rotating less faster than red parts, so the disc is collapsing alongside the filaments at the transition between the two zones.
Discrepancies of the azimuthal speed are less important for high values of $\beta$. Distances are in code units.}
\label{fig:vphi}

\end{figure*}

In the previous part, we have established that in self-gravitating $\beta$-cooled discs, the gas condenses in mainly rotationally supported filaments.
The $\Sigma$-PDF is however not the direct result of the radial column density profile of the filament. 
Indeed, to reproduce a PDF with a slope $s$, very shallow power-law column density profile of slope $1/s$ would be required. 
Such slope are not measured in our simulations (Figure~\ref{fig:fil_prof}), not yielded by the analytical model (Figure~\ref{fig:fil_eq}) and would require unrealistic broad filament to reproduce all the dynamic range of the PDF.
The power-law slope is built by the filament and is not entirely set by their radial profile, so it is set alongside the filaments. 
Figure~\ref{fig:segmentation} confirms this assertion as it shows that the filaments are not equally dense along their spine.
This reveals that there is a crucial second step to explain the $\Sigma$-PDF. 
Once the gas is condensed into filaments, it continues to collapse but along the filament ridge. 
To visualize collapsing motion,
Figure~\ref{fig:vphi} displays a bi-dimensional
map of the excess of tangential velocity
$v_\varphi/ v_{\mathrm{kepl}} -1$. High gradients 
are clearly visible and reveal ongoing collapse motions.
The filaments are over-Keplerian at some places and under-Keplerian at others, meaning that the gas accumulates at the transition. 
In fragmented discs, the fragments coincide with these accumulation zones. 
Obviously the collapse is stronger, meaning 
that the velocity gradients are higher, when $\beta$ is lower. 
Qualitatively speaking, the fragmentation 
of the filamentary spiral pattern resembles
the fragmentation of thermally 
supported self-gravitating filaments 
as studied for instance by 
\cite{1964ApJ...140.1056O} and \cite{2000MNRAS.311..105F}.
The most important difference is that the 
filamentary spiral pattern is radially supported 
by the differential rotation.

The idea of a two-step scenario (condensation within spiral arm and then fragmentation) was previously proposed by \cite{2016MNRAS.458.3597T}. They provide a fragmentation criterion within the filaments: a filament fragments if the value of the Toomre $Q$ parameter within it is below $1$. 
Our maps of the Toomre $Q$ parameter within our simulations (see Figure \ref{fig:Qmap} in the appendix) shows that $Q < 0.6$ is indeed a necessary condition for fragmentation.
Our work on the characterization of the $\Sigma$-PDF in section \ref{sec:result} gives an additional quantification of the amount of fragments expected to be found in the filaments, which is not given by the $Q < 0.6$ criterion. However, our computation of the probability to forms fragments needs to be generalized to take into account the effect of irradiation and more realistic cooling.

\section{Discussions and Conclusions}
\label{sec:conclusions}

We have presented simulations of self-gravitating disc undergoing a simple model of cooling, the $\beta$-cooling (section~\ref{sec:simu}). Using the Godunov scheme implemented in \textsc{Ramses}, we found a value of the fragmentation limit $\beta_{\mathrm{crit}}$ around 9, in rough accordance to previous results from simulations using SPH (section \ref{subsec:boundary}). 

However, the fragmentation limit is quite blurry, both in our simulations (it is not converged and is sensible to the definition we choose) and in the literature as shown in the introduction.
We found that the tendency of a disc to form fragments is better described by the probability density function of the fluctuation of the column density ($\Sigma$-PDF, section \ref{subsec:pdf}) as the PDF is more flat for strong cooling (low value of $\beta$), matching the excess of fragmentation (section~\ref{subsec:sbeta}). 
Actually, the slope of the PDF depends linearly on $\beta$ and this linear dependence can be understood using the relation between $\beta$ and the turbulent parameter $\alpha$ found by \cite{2001ApJ...553..174G}. The evolution of the slope of the PDF is thus the result of a balance of energy (section \ref{subsec:sbeta_model}).
In some extent, our conclusions are similar to those of \cite{2011MNRAS.416L..65P}, as we found that is no such thing as a clear fragmentation boundary, but the probability of forming bound fragments diminish as $\beta$ increases. 
The formula \eqref{eq:pfrag} of section~\ref{subsec:predictive} is a first attempt to empirically capture this behaviour, but it ought to be better constrained with simulations with higher $\beta$.
We tried to better understand the process leading to the formation of the PDF and fragmentation. From our study, we propose a two-step scenario:
\begin{enumerate}
    \item First, the gas form radially stable, rotationally supported filaments. The rotational support is stronger for less efficient cooling, as the angular momentum is less efficiently dissipated (section \ref{sec:stepone}).  
    \item Secondly, the gas collapses alongside the filaments to form the dense part of the PDF and eventually fragments (section \ref{sec:steptwo}).
\end{enumerate}

The investigation of what may first look like a simple academic problem lead us towards a better qualitative understanding of the process of fragmentation in self-gravitating discs.
A further improvement of this work would be to have a more quantitative view of the whole process, and especially the second step of our scenario that we have just brushed here. 
Simulation taking into account the irradiation of the stars shows it can efficiently suppress fragmentation, except far away from the star \citep{2011MNRAS.418.1356R, 2012ApJ...746..110Z}. Another interesting question would be to know to what extent the scenario for fragmentation remains valid in precence of irradiation and with a more realistic cooling function.

\section*{Acknowledgements}

We thank the anonymous referee for their useful comments that helped to improve the article. Moreover we thank Damien Chapon for his work on the \textsc{pymses} library that greatly helps the analysis of \textsc{ramses} outputs, and on the Galactica Database. We also thank Pierre Kestener for his support at the beginning of the project and his help with the code \textsc{CanoP}.
This work was granted access to HPC resources on the CINES's Occigen supercomputer under the allocation DARI A0050407023.

\section*{Software}

We made use of the following software and analysis tools:
GNU/Linux, \textsc{ramses} \citep{2002A&A...385..337T}, \textsc{python}, \textsc{Matplotlib} \citep{Hunter:2007}, \textsc{numpy} \citep{van2011numpy}, 
\textsc{pymses}, \textsc{Astrophysix}, \textsc{canop},
\textsc{astropy} \citep{astropy:2013, astropy:2018}, \textsc{filfinder} \citep{2015MNRAS.452.3435K}, 
\textsc{radfil} \citep{2018ApJ...864..152Z} and \textsc{idesolver} \citep{Karpel2018}. Thanks to the authors for making them publicly available.

\section*{Data availability}

The data underlying this article are available in the Galactica Database at \url{http://www.galactica-simulations.eu}, and can be accessed with the unique identifier \textsc{\href{http://www.galactica-simulations.eu/db/STAR_PLANET_INT/FRAGDISK}{FRAGDISK}}.
Additional data and the source code used to run simulations and perform analysis will be shared on reasonable request to the corresponding author.




\bibliographystyle{mnras}
\bibliography{bibliography/frag}


\appendix

\section{Conservation of angular momentum}

\begin{figure}
    \centering
    \includegraphics[width = 0.5 \textwidth]{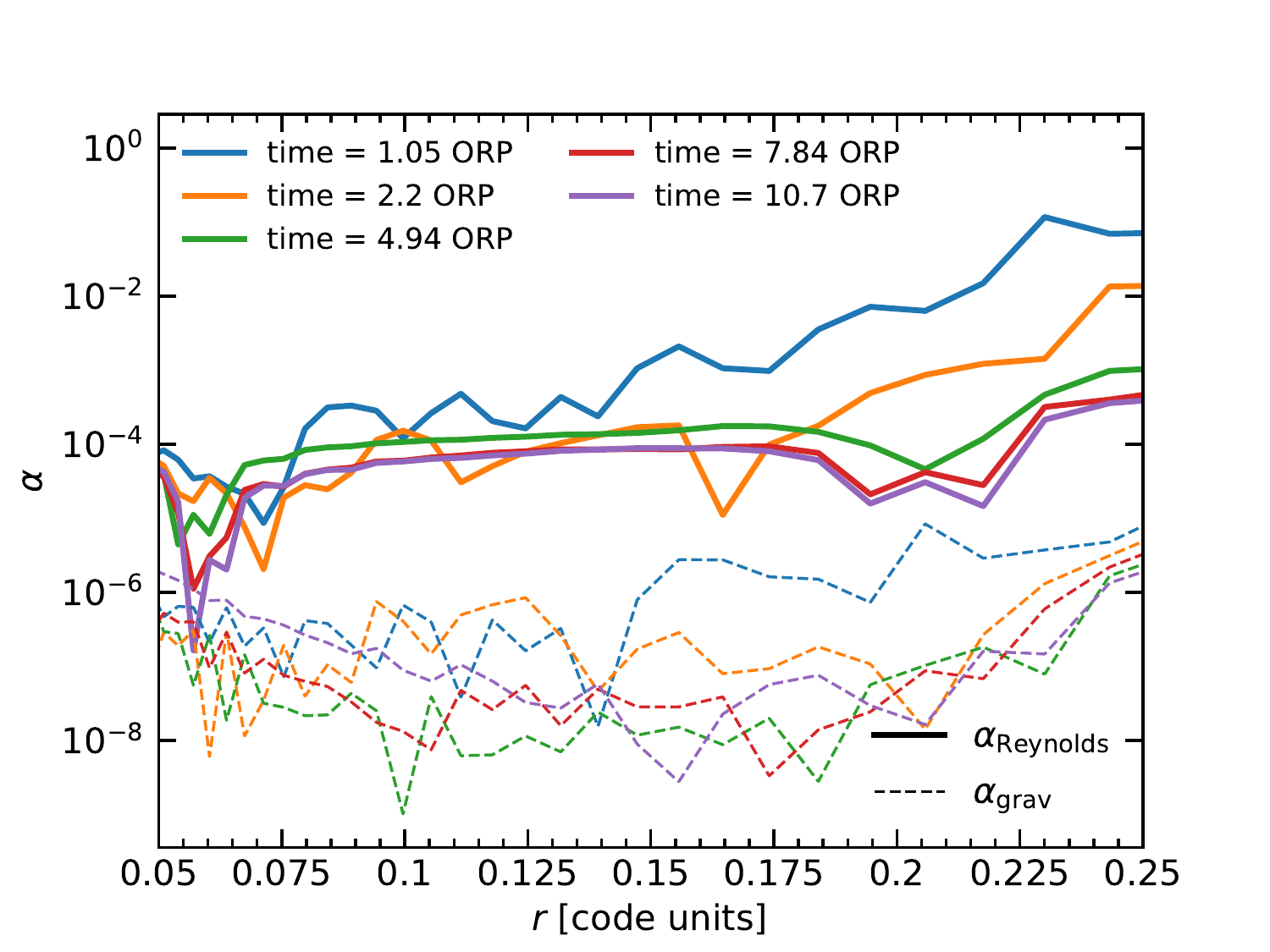}
    \caption{Value of $\alpha$ in a stable non-cooled disc with $Q \geq 3$.}
    \label{fig:alphaQ3}
\end{figure}

As mentioned in section \ref{sec:simu}, the poor conservation of the total angular momentum can be a concern when using a Cartesian grid for an cylindrical problem \citep{2015A&A...579A..32L}. We run a simulation of a stable disc to repeat the measurement done by  \cite{2017A&A...599A..86H} in their Figure 2. The goal is measure how the bad conservation of angular momentum can influence the fragmentation by artificially increasing the turbulence in the disc. 
The set-up we use is the same as described in section \ref{sec:simu}, except that the temperature is initially set so that the value of $Q$ at the outer edge of the disc is 3, and the $\beta$-cooling is switched off. If the conservation of angular momentum was perfect, we would expect that the disc remains completely stable and that the value of the Shakura \& Sunayev’s $\alpha$ parameter remains equal to $0$ or at least very close. 
Departure from $\alpha = 0$ can be seen as an artificial stress due to various numerical effects, including a bad conservation of angular momentum.
Figure~\ref{fig:alphaQ3} show the contribution of the Reynolds and gravitational stress tensors to the parameter $\alpha$, respectively $\alpha_{\mathrm{Reynolds}}$ and $\alpha_{\mathrm{grav}}$, computed as in \cite{2004MNRAS.351..630L}. The value of $\alpha$ for this artificial stress goes quickly below $10^{-4}$ and thus has not any significant impact on the fragmentation process.

\label{sec:consangmom}

\section{Toomre's Q parameter}

\begin{figure*}
    \centering
    \includegraphics[width = 0.6 \textwidth]{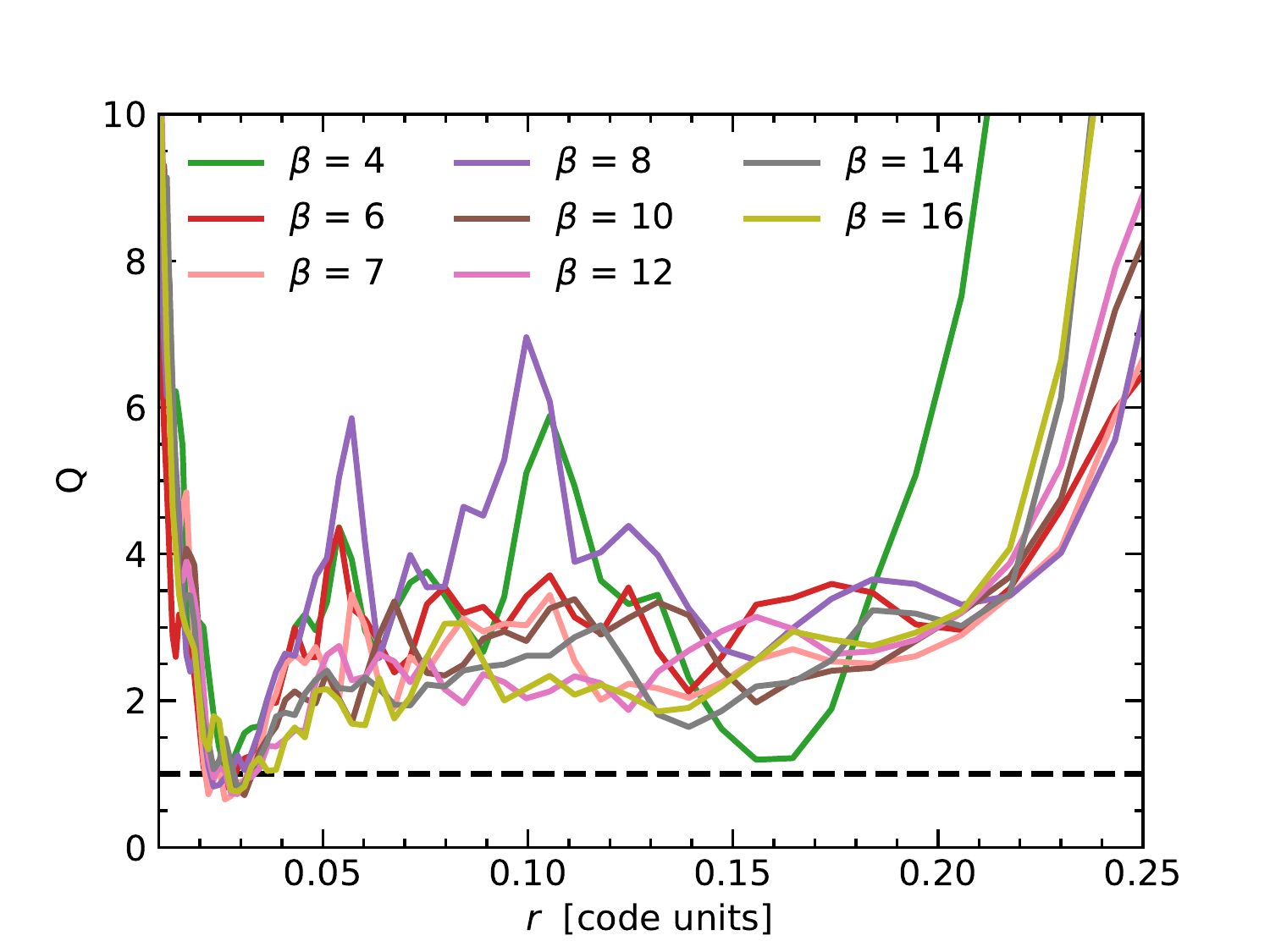}
    \caption{Radial profile of the Q Toomre parameter for the \textsc{JR13\_TIC} simulation, from snapshots at the same time as Figure \ref{fig:coldens_jr13}. The black dashed line corresponds to the critical value $Q = 1$.}
    \label{fig:Q}
\end{figure*}

\begin{figure*}
\centering
\includegraphics[width=\textwidth]{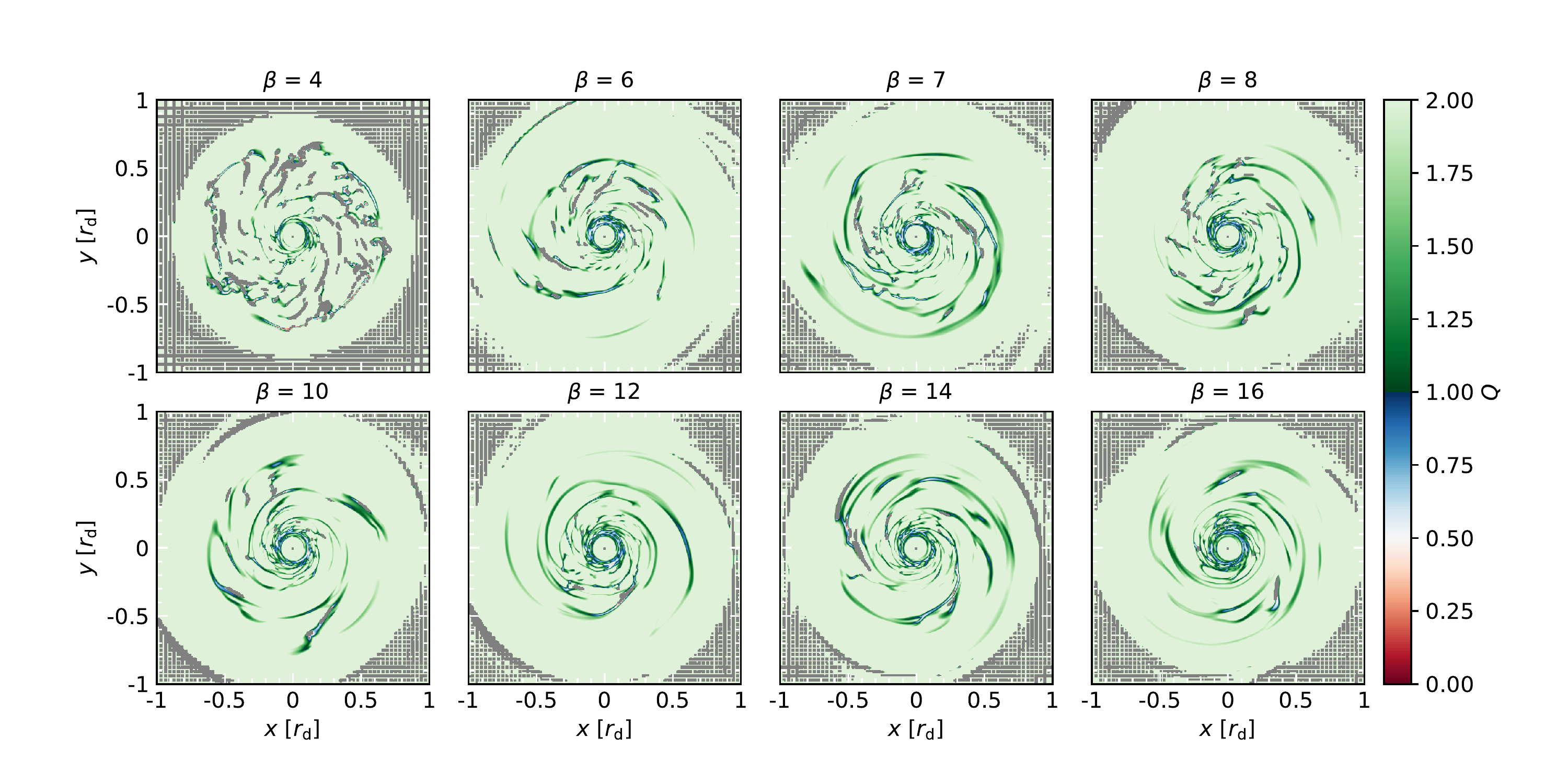} 

\caption{Map of the Toomre Q parameter in the disc. Zones with $Q < 0.6$ appear in white or red and are located in the filaments. Green zones are gravitational stable according to the Toomre criterion. In the grey zones the epicyclic frequency was not computed because of the adaptive resolution.}

\label{fig:Qmap}

\end{figure*}

The Toomre $Q$ parameter \citep{1964ApJ...139.1217T}
 \begin{equation}
\label{eq:toomre_apx}
Q = \dfrac{c_s \kappa}{\pi G \Sigma}.
\end{equation}
is crucial to quantify the stability of self-gravitating disc \citep{2001ApJ...553..174G,2016MNRAS.458.3597T}. Figure \ref{fig:Q} features a figure of the azimuthally averaged value of $Q$ in the simulations from the group \textsc{JR13\_TIC}. For $\beta > 10$ the radial profile of $Q$ is a plateau with $Q \approx 2\textrm{--}3$. 
Note that this is the commonly accepted critical value for gravitational instability in presence of non-axisymmetric perturbations \citep[and references therein]{2017MNRAS.469..286R}. 
Lower values of $\beta$ yield higher values of the azimuthal average of the $Q$ parameter, which may be found surprising but is explained by the fact the in these simulations, the mass in concentrated in very thin filaments. Indeed, Figure \ref{fig:Qmap} featuring the map of $Q$ shows that in these simulations $Q$ is very low in the fragmenting filaments and quite high in the rest of the disc.
By comparing Figure \ref{fig:Qmap} and Figure \ref{fig:coldens_jr13}, we can notice that, as stated by \cite{2016MNRAS.458.3597T}, $Q < 0.6$ is a necessary condition for fragmentation within the filaments, but we also have non-fragmenting filaments with $Q < 0.6$, in the $\beta = 16$ simulation for instance.

\bsp	
\label{lastpage}
\end{document}